\documentclass[fleqn,usenatbib]{mnras}

\usepackage{txfonts}

\usepackage[T1]{fontenc}

\DeclareRobustCommand{\VAN}[3]{#2}
\let\VANthebibliography\thebibliography
\def\thebibliography{\DeclareRobustCommand{\VAN}[3]{##3}\VANthebibliography}

\usepackage{graphicx}	
\usepackage{subcaption}
\usepackage{lscape}	
\usepackage{float}
\usepackage{setspace}
\usepackage{threeparttable}
\usepackage{placeins}   
\usepackage[figuresright]{rotating}   
\usepackage{pdflscape}   

\newcommand{\xsp}{XSPECT\,}	
\newcommand{\fouru}{4U~1608$-$52\,}	
\newcommand{\aql}{Aql~X$-1$\,}	
\usepackage{multirow}
\usepackage{verbatim}
\usepackage{enumitem}   
\usepackage{calc}       

\title[SXRTs with \xsp]{Tracing the outburst decay of soft X-ray transients \aql and \fouru with XSPECT}

\author[Chatterjee et al.]{
Rwitika Chatterjee,$^{1}$\thanks{E-mail: rwitika@ursc.gov.in}
Vivek Kumar Agrawal,$^{1}$
V. P. Shyam Prakash,$^{1}$
Koushal Vadodariya,$^{1}$
and Radhakrishna V$^{1}$
\\
$^{1}$Space Astronomy Group, ISITE Campus, U. R. Rao Satellite Centre, ISRO, Bengaluru 560037, India\\
}

\date{Accepted XXX. Received YYY; in original form ZZZ}

\pubyear{2026}

\begin{document}
\label{firstpage}
\pagerange{\pageref{firstpage}--\pageref{lastpage}}
\maketitle

\begin{abstract}
\xsp instrument on-board XPoSat mission is a soft X-ray spectrometer sensitive in the energy band 0.8$-$15~keV. \xsp has observed several bright neutron star low mass X-ray binaries since launch. Two well known sources, \aql and \fouru which are soft X-ray transients, were observed by XPoSat during the decay phase of their recent outbursts in September~2024 and February~2025 respectively. During \xsp observations, \fouru exhibited a superburst which is a long duration thermonuclear burst, believed to be triggered by carbon burning. We carry out a detailed spectro-temporal analysis of the superburst, tracing its onset, rise, and decay over the next several hours. Using time-resolved spectroscopy, we probe the spectral evolution of the source and find that the persistent emission is suppressed during the superburst and the emission can be described by a gradually cooling blackbody component. The superburst was preceded by a precursor burst which is a normal type-I X-ray burst. We also observe a type-I burst $\sim 5$~days after the superburst, indicating resumption of burst activities which is typically quenched after a superburst. \aql also exhibited a type-I burst during \xsp observations. The persistent emission of both the sources can be fitted using a combination of blackbody and disk blackbody emission or, alternatively, using a disk Comptonized by an optically thick plasma. Using the latter model, we find a clear flux dependence of the Comptonization parameters, with both the sources exhibiting harder spectra at higher accretion rates. 
\end{abstract}

\begin{keywords}
accretion, accretion discs -- X-rays: binaries -- X-rays: bursts -- X-rays: individual: \aql, \fouru
\end{keywords}



\section{Introduction}
\label{sec:intro}

XPoSat is, after AstroSat \citep{singh2014}, India's second mission dedicated to astronomy, and the first mission aimed to carry out X-ray spectro-polarimetric studies. It consists of two co-aligned payloads: The Polarimeter Instrument in X-rays (POLIX, \citealt{paul2024}), a thomson scattering based medium energy ($8-30$~keV) polarimeter, and the X-ray Spectroscopy and Timing (\xsp) instrument \citep{rkrish2025}, a soft X-ray ($0.8-15$~keV) spectrometer. XPoSat was launched on 2024 January 1 from Sriharikota, India, and after completing its performance verification (PV) phase, commenced its science operations from early March 2024. A description of the ground tests and on-board calibration of \xsp during the PV phase, demonstrating the instrument capability, can be found in \cite{chatterjee2025}.

Since polarimetry is a photon-hungry technique, the mission is designed to make long-term observations of bright sources. Taking advantage of this requirement, \xsp has a unique capability to carry out detailed soft X-ray spectral and timing studies of the sources being observed over several days to weeks. \xsp is a collimated X-ray telescope, which uses second generation Swept Charge Devices (SCDs, \citealt{lowe2001, holland2008}) as the detectors. The instrument employs two kinds of co-aligned square collimators, with $\sim 2^{\circ}\times 2^{\circ}$ and $\sim 3^{\circ}\times 3^{\circ}$ edge-to-edge fields of view (FOV). Even though the FOVs are significantly large, the sources are chosen such that there are no other bright X-ray sources within the FOV above the sensitivity level causing source confusion. 

One of the prime science targets of \xsp observations are bright neutron star low mass X-ray binaries (NS-LMXBs). NS-LMXBs are binary systems in which a weakly magnetized neutron star accretes matter from a low mass ($\lesssim 1$~M$_{\odot}$) companion, primarily via Roche lobe overflow. Depending on the pattern traced by their emission in the color-color diagram (CCD) and hardness-intensity diagram (HID), these sources are divided into two categories: `Z' and `atoll' sources \citep{hasinger1989}. The Z and atoll sources have different spectro-temporal properties. Z sources are characterised by a high mass accretion rate, and high persistent luminosities, with the Z-track tracing the horizontal, normal and flaring branches. On the other hand, atoll sources typically have a much larger range of luminosity variation, with the sources occupying the (softer) banana state and (harder) island states.

The spectra of NS-LMXBs typically consist of two main components - thermal emission and Comptonized emission. Traditionally, two different scenarios have been proposed to explain the origins of these two components. In the `eastern' model (\citealt{mitsuda1989, disalvo2002, agrawal2003}), the thermal emission arises from the accretion disk, modeled by a multi-color blackbody component, which is directly observed, along with Comptonized photons arising from a Comptonized blackbody component at the boundary layer. The `western' model (\citealt{white1988, disalvo2000}) proposes a single color blackbody emission from the hot neutron star surface or boundary layer, with the thermal disk photons upscattered by an accretion disk Corona. In practice, different combinations of thermal and Comptonized components can provide statistically comparable descriptions of NS-LMXB spectra, despite implying different physical origins for the emission, highlighting the ineherent degeneracy in spectral decomposition of these sources. In addition to these two components, reflection features are also often present in their spectra, in the form of iron line emission, Compton back-scattering hump etc (\citealt{fabian1989, ross1999}).

NS-LMXBs are known to exhibit a wide range of X-ray variability phenomena, including quasi-periodic oscillations (QPOs) spanning mHz to kHz frequencies, and broadband noise features in their power density spectra \citep{vanderklis2006}. They also often exhibit thermonuclear (type-I) X-ray bursts, attributed to unstable burning of accreted material, such as hydrogen and/or helium, on the surface of the NS (\citealt{lewin1993, strohmayer2006}). The duration of most type-I bursts range from a few seconds to about a minute, with the light curve following a fast rise (flash) followed by an exponential decay (NS cooling) profile. 

In addition to type-I bursts, some of these sources have also exhibited `superbursts' (see, e.g., \citealt{intzand2017, alizai2023} for reviews), which are similar to type-I bursts in terms of temporal profile, but have much longer decay times of the order of several hours. These are rare events with only $\sim 30$ detected from 16 sources till date. Superbursts are believed to be due to unstable Carbon fusion in the H/He ashes, occuring at much larger depths, and releasing few orders of magnitude more energy than the common bursts (\citealt{cumming2001, strohmayer2002}). Their long durations and high fluence imply that they can significantly perturb the surrounding accretion environment, affecting the inner disc and persistent emission. In many of the cases, `precursor' bursts which are similar to normal type-I bursts are observed before the superburst. Typically, superbursts are detected in systems accreting at $\sim 0.1-0.25$ times the Eddington limit, but the luminous Z-source GX~$17+2$, accreting at close to Eddington rates, has also shown superbursts \citep{intzand2004}. The peak flux of most superbursts are sub-Eddington. However a superburst from 4U~$1820-30$, an ultracompact binary with a likely hydrogen-deficient dwarf companion, reached the Eddington limit \citep{strohmayer2002}. Superburst spectra are generally modelled with a blackbody component, similar to type-I bursts, which cools as the superburst decays (e.g. \citealt{strohmayer2002, cornelisse2002, keek2014}).

Being bright sources in the soft X-ray band having a rich variety of observational phenomenology, \xsp is particularly well-suited to study NS-LMXBs. \xsp can carry out pile-up free observations of these bright sources, and long duration observations make this an ideal platform to study spectro-temporal evolution and state changes. In this paper, we present a detailed spectral and temporal analysis of two sources, \aql and \fouru, which were observed by \xsp during their recent outbursts (September~2024 and February~2025 respectively). During the observations, we have observed two type-I bursts from \fouru and one from \aql. In addition, we also observed a superburst from \fouru. We have studied the nature of persistent, as well as burst and superburst emission. 

These two sources are well-known and well-studied soft X-ray transients (SXRTs, see \citealt{campana1998} for a review). Both of these are atoll sources, and show recurrent outbursts when their luminosity increases by several orders of magnitude. Typically, during the outbursts, the source is in the high soft state. This state is dominated by thermal emission, modelled either as a single or multi color blackbody. On the other hand, the quiescent (low hard) state is characterised by harder spectra, with a dominant Comptonized component. Type-I X-ray bursts have been observed in the active phase of both the sources. 

The rest of the manuscript is organized as follows: Sections~\ref{subsec:introaql} and \ref{subsec:intro4u} provide brief introduction on the two sources, Section~\ref{sec:obslis} describes the observations and data reduction methods, Section~\ref{sec:res} discusses the spectral and temporal analysis as well as the results, and Section~\ref{sec:disc} concludes the paper with a short discussion.

\subsection{Aql~X$\mathbf{-1}$}
\label{subsec:introaql}
\aql , discovered in 1965 \citep{friedman1967}, is an NS-LMXB, with a K-type main sequence star as the companion. It is one of the most active known SXRT, which regularly goes into outburst roughly once per year (\citealt{simon2002, gungor2014}). The outbursts of \aql are typically characterised by a fast rise over $5-10$~days, followed by a slow, nearly exponential decays with a e-folding time of $30-70$~days \citep{campana1998}. The system has an orbital period of $\sim 18.9$~hr \citep{chevalier1991}. \cite{matasanchez2017} use NIR spectroscopy to constrain the source to be located at a distance of $6\pm 2$~kpc with an orbital inclination between 36$^{\circ}$ and 47$^{\circ}$. Using the peak flux of PRE bursts, the distance is estimated as $5.0\pm 0.9$~kpc \citep{galloway2008}, which we adopt in this paper.

Intermittent coherent pulsations at 550.27~Hz were detected from this source by \cite{casella2008} during the rising phase of its 1998 outburst. In addition, burst oscillations were also identified from this source at $\sim 549$~Hz \citep{zhang1998}. The power spectrum of \aql shows mHz (\citealt{revnivtsev2001, mancuso2021}), low frequency \citealt{yu2003, zhang2015} as well as kilohertz QPOs (\citealt{cui1998, mendez2001, barret2008}). 

This source has shown several type-I X-ray bursts, which are extensively studied in literature (\citealt{koyama1981, galloway2008, keek2018, guver2022}). In addition, two superbursts from this source have been detected by MAXI (\citealt{serino2016, iwakiri2020}), one of which was also partially observed by NICER \citep{li2021}.

\subsection{4U~$\mathbf{1608-52}$}
\label{subsec:intro4u}
\fouru is a transient atoll source discovered in 1972 by the Vela~5 satellites \citep{belian1976}, and subsequently confirmed by Uhuru observations \citep{tananbaum1976}. It is located in the Norma constellation, with QX~Normae as its optical counterpart \citep{grindlay1978}. Periodic modulations in I-band data suggest an orbital period of 12.9~hr \citep{wachter2002}. The estimated distance to this source, assuming solar composition, is $3.2\pm 0.3$~kpc based on the peak flux of PRE bursts \citep{galloway2008}, whereas \cite{poutanen2014} obtain a range of $3.1-3.7$~kpc using hard state bursts, and comparing with theoretical models. \cite{guver2010} find a best-fit distance of $5.8^{+2.0}_{-1.9}$~kpc, whereas \cite{ozel2016} quote a minimum distance of 3~kpc with a most likely distance of 4~kpc. In this paper, we adopt a distance of $3.2\pm 0.3$~kpc to the source, which is further discussed in Section~\ref{subsec:spec_bur}. \cite{degenaar2015} modeled the NuSTAR spectra of this source which shows evidence of disk reflection, with a binary inclination of $\simeq 30^{\circ}-40^{\circ}$ -- here, we adopt a value of $35^{\circ}$.

Sources like \aql typically show a fast rise and exponential decay profile in their outburst light curves. There is another category of SXRTs which shows extended, alternating periods of relatively high and low fluxes. The properties of \fouru are intermediate between these two classes (\citealt{lochner1994, campana1998}), and it undergoes recurrent outbursts every $1-2$~yr. It is also a prolific X-ray burster (e.g. \citealt{jaisawal2019, guver2021, chen2022}), with about 145 recorded bursts in the MINBAR sample \citep{galloway2020}. Burst oscillations during several PRE (photospheric radius expansion) bursts from this source has revealed a spin frequency of 619~Hz \citep{hartman2003}, making this one of the fastest rotating NS. Two superbursts have also been observed from this source till date - one by RXTE/ASM \citep{keek2008} and the other by MAXI (see footnote in \citealt{alizai2023}). 

This source, like \aql, has also shown mHz QPOs (\citealt{revnivtsev2001, mancuso2021}) as well as kHz QPOs (\citealt{berger1996, yu1997, mendez1998}). Low frequency QPO and noise components are also seen in its power density spectra (\citealt{yoshida1993, yu1997}).

\section{Observations and Data Reduction}
\label{sec:obslis}

\begin{table*}
\centering
\caption{XSPECT observation log of Aql X-1 and 4U $1608-52$}
\label{tab:obslog}
\begin{tabular}{ccccccc}
\hline
Source                      & Proposal ID                & Segment & Start date  & \multicolumn{1}{c}{Start MJD}  & \multicolumn{1}{c}{End MJD} & On-source exposure (ks) \\ \hline
Aql X-1                     & T24\_0009                  & 1       & 2024 Oct 4  & \multicolumn{1}{c}{60587.0044} & 60594.0270                  & 113.6                   \\
\multirow{2}{*}{4U 1608-52} & \multirow{2}{*}{T25\_0004} & 1       & 2025 Mar 8  & 60742.0704                     & 60744.9441                  & 43.4                    \\
                            &                            & 2       & 2025 Mar 17 & 60751.1328                     & 60761.0000                  & 208.5                   \\ \hline
\end{tabular}
\end{table*}

\begin{figure*}
\centering
\begin{subfigure}{0.49\textwidth}
\includegraphics[scale=0.25, trim=2cm 1.cm 1cm 1.5cm, clip=true]{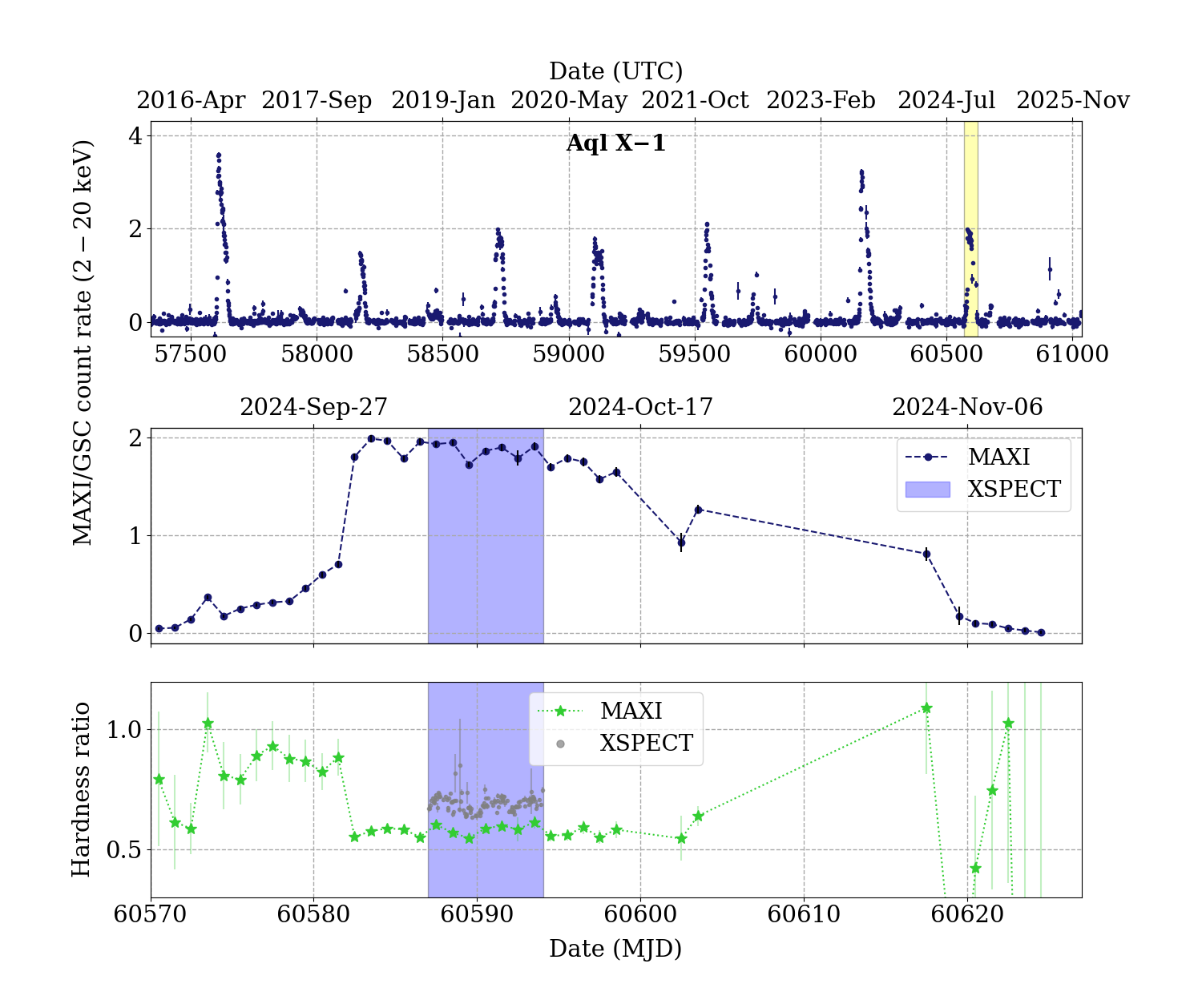}
\caption{}
\end{subfigure}\hfill
\begin{subfigure}{0.49\textwidth}
\includegraphics[scale=0.25, trim=2cm 1.cm 1cm 1.5cm, clip=true]{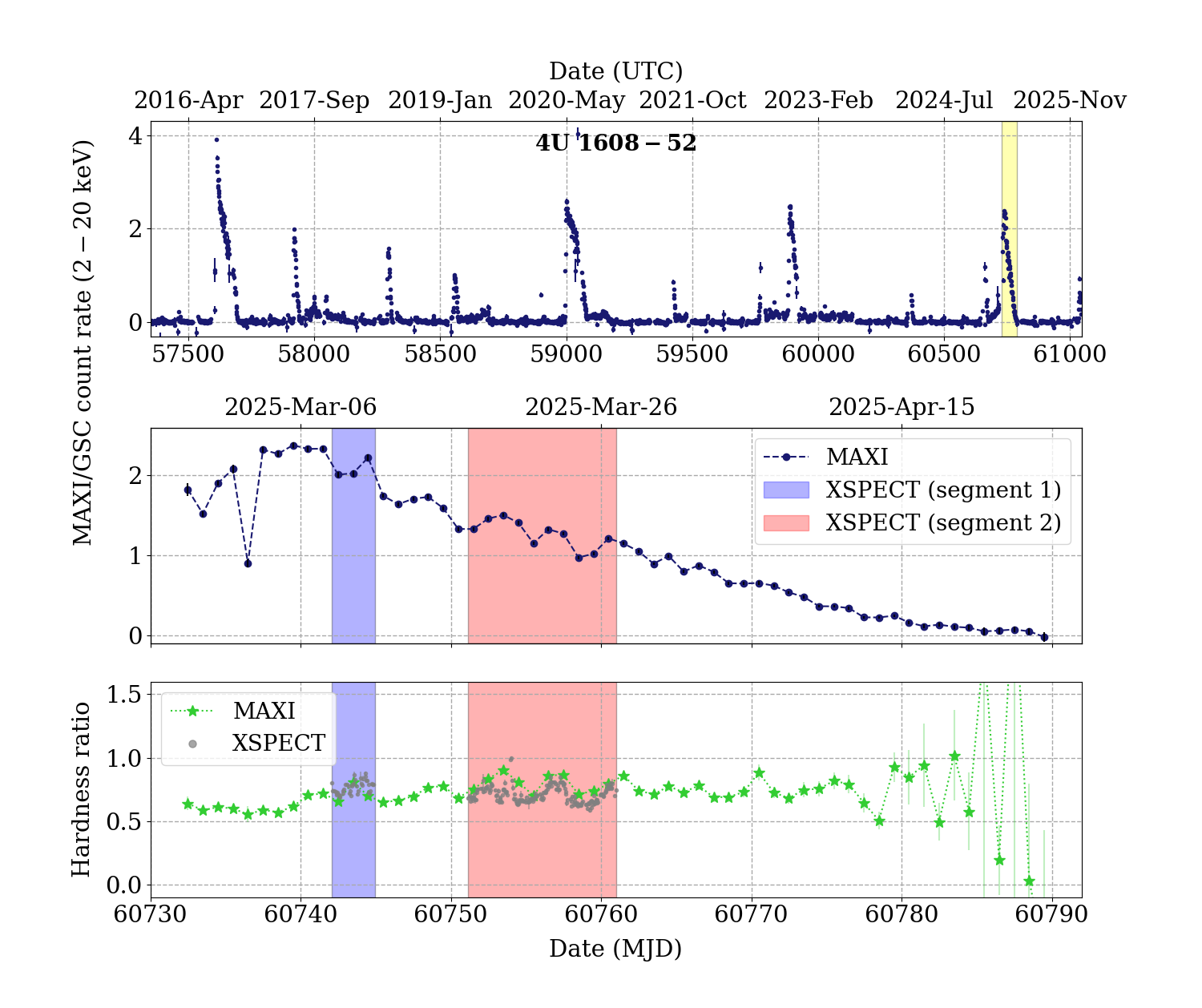}
\caption{}
\end{subfigure}  
\caption{(Top panel) MAXI/GSC $2-20$~keV long-term light curve of (a) \aql, and (b) \fouru. (Middle panel) Zoomed in portion of the same light curve corresponding to the shaded portion in yellow in the top panel, covering the outburst during which \xsp observations were made. The \xsp observation segments are marked. (Bottom panel) MAXI and \xsp hardness ratio ($4-10$~keV/$2-4$~keV) evolution during the outburst. }
\label{fig:maxiaql4u}
\end{figure*} 

We triggered \xsp Target of Opportunity (ToO) observations during the latest outburst of these two sources. \xsp observed \aql during 2024~Oct~$4-10$, tracing the initial decay phase of the outburst, just beyond the outburst peak. \fouru was observed during 2025~Mar~$8-10$, and again during 2025~Mar~$17-26$. The observation log of both the sources is summarized in Table~\ref{tab:obslog}. The long-term MAXI/GSC light curve of the two sources is shown in Figure~\ref{fig:maxiaql4u}, with the \xsp observation epochs marked. The hardness ratio, defined as the ratio of counts in $4-10$~keV to $2-4$~keV, is also plotted in the bottom panel with green markers. For comparison, the hardness ratio is also computed from \xsp data in the same energy bands, which is overplotted. There is a good match between the two instruments, and during the \xsp observations, the hardness ratio of both the sources does not show any significant variation.

\xsp data was analyzed using \texttt{xspect\_pipeline\_l1l2} software (v2.2.1)\footnote{XSPECT software is hosted at PRADAN: https://pradan1.issdc.gov.in/x01}. The level~1 data was screened using the nominal filtering criteria, as described in \cite{chatterjee2025}, using \textit{xspl2screen} to generate the screened (level~2) event files. The screening applies filtering on the data to retain events only in the good time intervals. We checked for the occurence of particle flares using \textit{xspfilterflares}. Such high energy events were present in few orbits of each day of \aql observations, and these durations were also excluded from further analysis. Subsequently, we merged the event files of each segment\footnote{Segment definition in Table~\ref{tab:obslog}} using \textit{xspevtmerge}.

From the segment-wise merged event files, we first generated detector-wise light curves in the full energy band ($0.8-15$~keV). For the case of \fouru, it was evident from the light curves that SCD~4 (and SCDs~$5-7$ to a lesser extent) shows some signs of optical light contamination in the data, leading to anomalously higher counts than the other detectors. These four SCDs belong to the same `quad' module \citep{rkrish2025}, sharing the same block of memory in the data packets. Consequently, one SCD in a quad recording more events may lead to lower available space for the others, which can affect the reliability of the overall count rate estimate. However, note that these signatures are only seen near eclipse ingress or egress (i.e. when the spacecraft is not in complete shadow of the Earth), but never in full eclipse. For having uniform GTIs across all SCDs, we removed these 4 SCDs from our subsequent analysis and used only the remaining 11. For the case of \aql, we used data from all 15 SCDs, as no such excess counts were observed\footnote{Since launch, this has been observed in a handful of sources, and it depends on the exact spacecraft atitude and position in its orbit. Detailed analysis for understanding the same is underway.}. 

Lightcurves in different energy bands and with different binsizes were generated using \textit{xspl2lcgen} and spectra were created using \textit{xspl2specgen}. We used the `static' background in our spectral fits, which were generated by setting the option -{}-{}backspecgen to True in \textit{xspl2specgen}. Similarly, background light curves were generated by setting -{}-{}backlcgen to True in \textit{xspl2lcgen}. For both the sources, the average $0.8-15$~keV and $0.8-11$~keV background count rate over the observations was found to be 0.20~cps~SCD$^{-1}$ and 0.08~cps~SCD$^{-1}$ respectively. As explained in \cite{chatterjee2025}, we used only the single event light curves and spectra in our analysis. Detector selection was also specified while generating the light curves and spectra of \fouru, and appropriate effective area files (or ARFs) were generated by setting the option -{}-{}arfgen to True while running \textit{xspl2specgen}.

Since the instrument is sensitive in `soft' X-rays, having a limited energy band ($0.8-15$~keV), the choice of bands for hard color and soft color are critical. The shapes of the CCD and HID are highly sensitive to the choice of energy bands. To avoid this, we imposed certain restrictions while selecting the energy bands. We ensured that the lightcurves are not dominated by background in any of the four bands (two each for soft color and hard color). Moreover, we also chose sub-bands such that they have comparable count rates (i.e. color $\sim 1$), and also avoided choosing boundaries with a steep increase/decrease of count rates. This gave us a stable and reliable energy band definition.

After several iterations, we finally defined the `soft color' as the ratio of count rates in the energy bands $2.5-4.5$~keV and $1-2.5$~keV, and the `hard color' as the ratio of count rates in $6-11$~keV and $4.5-6$~keV. Intensity is defined as the $0.8-11$~keV\footnote{Being soft sources, the spectra is background-dominated beyond 11~keV} count rate (from all 15~SCDs for \aql and for 11~SCDs for \fouru). We binned the light curves in 500~s bins, and used the background subtracted light curves to generate the CCDs and HIDs.  

Spectral fitting was done in \textit{XSPEC} \citep{arnaud1996} version~12.14.1. We used the response and background bundles X01\_XPC\_rsp\_20250319 and X01\_XPC\_bkg\_20250519 respectively, produced by the \xsp team, in spectral fitting. The spectra were binned using the \textit{HEASoft} task \texttt{ftgrouppha} using the optimal binning scheme \citep{kaastra2016}, to have a minimum signal-to-noise ratio of 3 per bin. We fitted the spectra in the range $0.8-11$~keV. A systematic uncertainty of 1\% is applied to the spectra. Two \texttt{edge} components at $\sim 1.5$~keV and $\sim 1.8$~keV, corresponding to Al and Si absorption edges, were required to improve the residuals near these energies \citep{chatterjee2025}. In all the spectral fits, we simultaneously fitted the spectra of the two FOVs, tying the parameters across the two spectra. We included a multiplicative constant to account for the cross-calibration between the two FOVs, freezing it to 1 for $2^{\circ}\times 2^{\circ}$ and allowing it to vary for $3^{\circ}\times 3^{\circ}$. For all the fits, the best-fit value of this constant for $3^{\circ}\times 3^{\circ}$ spectra was found to lie in the range $1.09-1.1$. Gain fit was applied using the \texttt{gain fit} command during spectral fitting to account for gain correction. 

\section{Analysis and Results}
\label{sec:res}

\begin{figure*}
	\centering
	\includegraphics[scale=0.27, trim=0cm 0cm 0cm 0cm, clip=true]{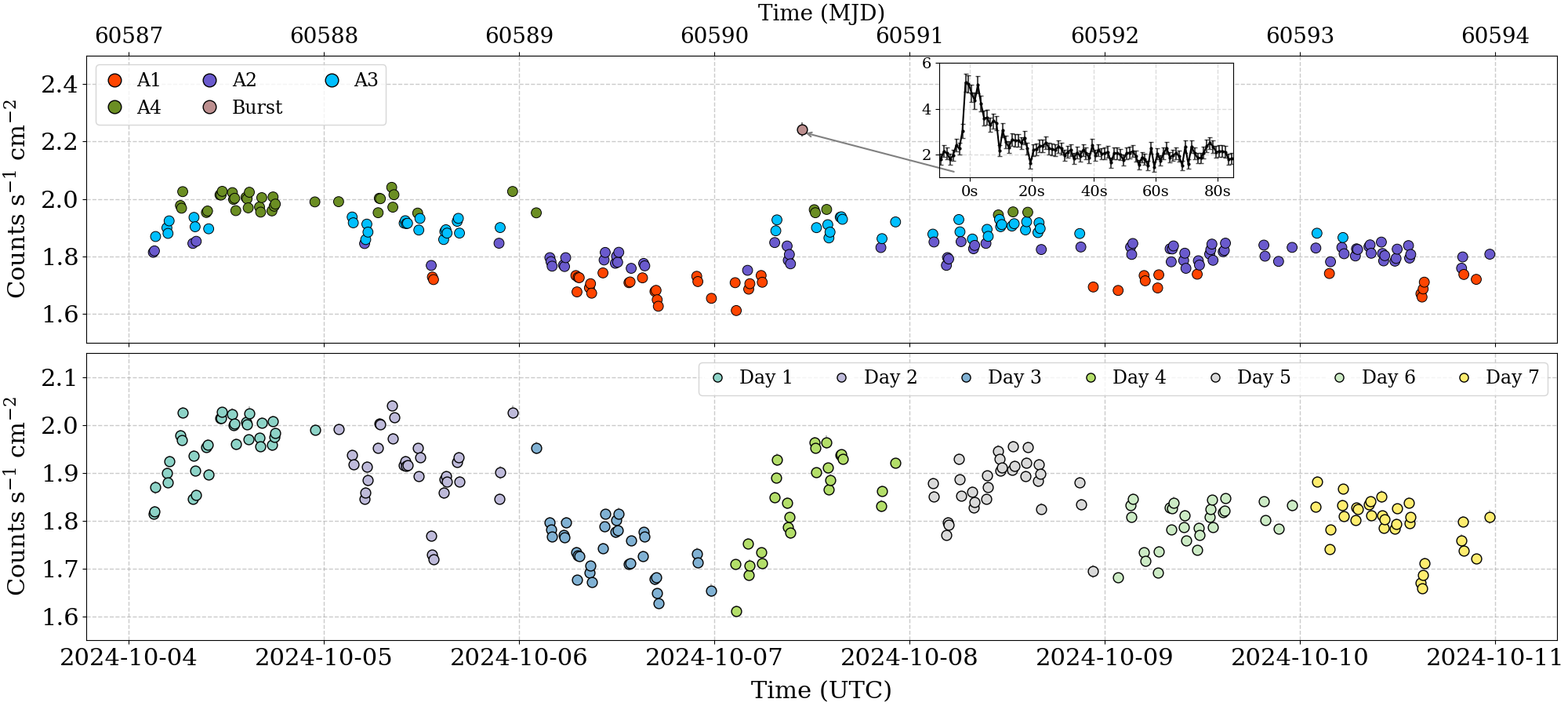}
    \caption{Background-subtracted $0.8-11$~keV light curve of \aql. Each point corresponds to 500~s. The count rate has been normalized by the number of detectors used (15 in this case). Inset shows a zoomed view of the type-I X-ray burst at 1~s binning. The markers in the top panel are color-coded by the region of HID they lie in. The bottom panel, showing only the persistent emission, are color-coded by day of observation.}
    \label{fig:fullaql}
\end{figure*}

\begin{figure*}
	\centering
	\includegraphics[scale=0.3, trim=0cm 0cm 0cm 0cm, clip=true]{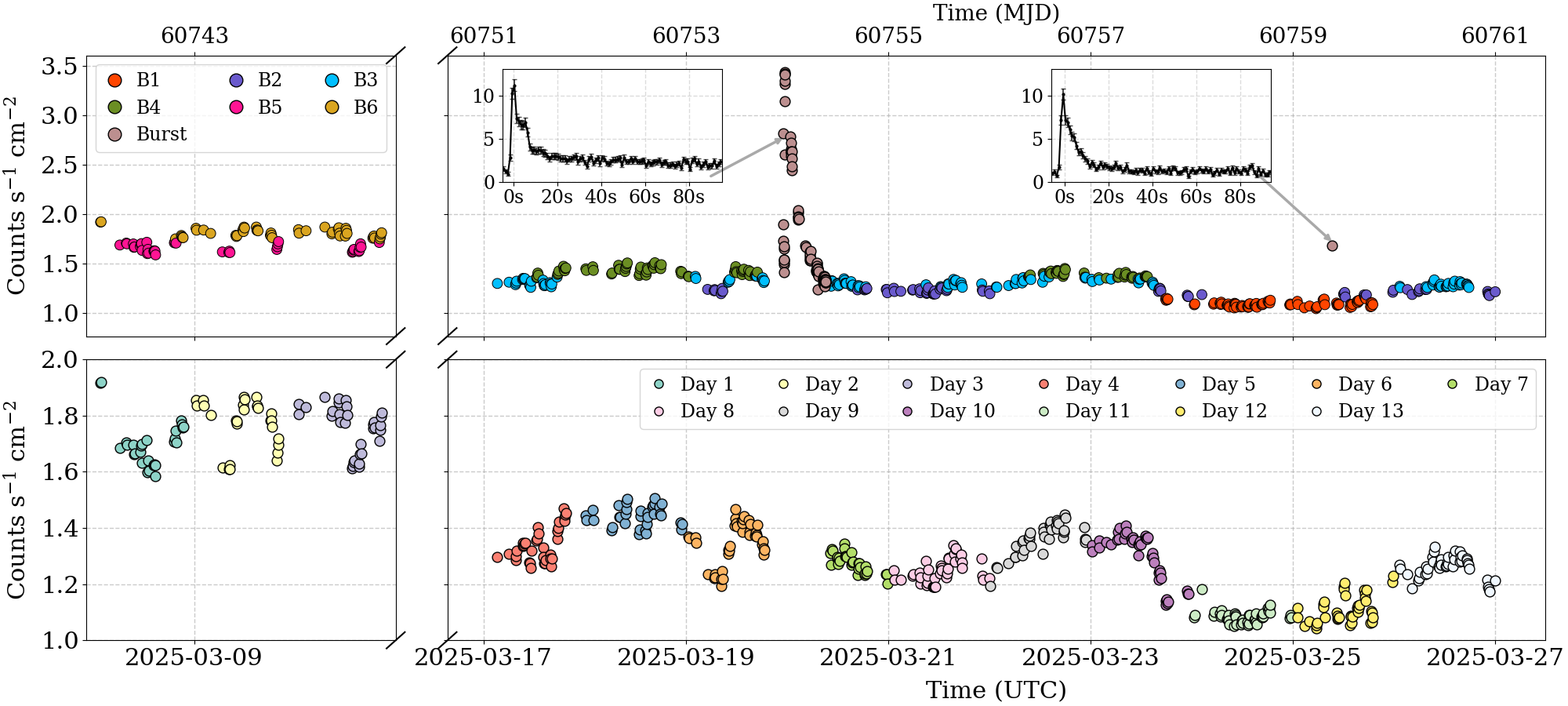}
    \caption{Background-subtracted $0.8-11$~keV light curve of \fouru. Each point corresponds to 500~s. The count rate has been normalized by the number of detectors used (11 in this case). Insets show a zoomed view of the two type-I X-ray bursts at 1~s binning The superburst, which occurred near the midnight of 2025~Mar~20 is clearly visible. Color-coding follows the same scheme as Figure~\ref{fig:fullaql}.}
    \label{fig:full4u}
\end{figure*}

The background-subtracted $0.8-11$~keV light curves, binned at 500~s, of both the sources are shown in Figures~\ref{fig:fullaql} and \ref{fig:full4u}. The light curve is plotted in counts~s$^{-1}$~cm$^{-2}$, with count rates normalized by the effective geometric collecting area of the selected detectors (accounting for detector area, collimator open-area fraction, and alignment corrections), to enable comparison independent of detector selection\footnote{Unless stated otherwise, count rates in subsequent figures and tables are reported in raw counts~s$^{-1}$ for the specified detector configuration.}. On the top panels, the persistent emission points are color-coded with their respective boxes in the HID (Section~\ref{subsec:tim_pers}), to follow the movement of the source across different regions during the observations. During the \xsp observations of \fouru, the source exhibited a superburst \citep{chatterjee2025atel}, which was also detected by MAXI \citep{serino2025}. This is the third known superburst from this source, the previous two being observed in May~2005 \citep{keek2008} and July~2020 \citep{boztepe2025}.

In addition to the superburst from \fouru, one and two thermonuclear Type-I X-ray bursts were detected from \aql and \fouru respectively, which are also shown in the insets in the top panel at 1~s binning. On the bottom panels of Figures~\ref{fig:fullaql} and \ref{fig:full4u}, the bursting durations are removed and only the persistent light curves are shown. Here, the markers are color-coded day-wise.

We carried out a detailed analysis of all the available observations of these two sources. In the following subsections, the timing (Section~\ref{subsec:tim}) and spectral (Section~\ref{subsec:spec}) analysis of these two sources are described, along with the results obtained.

\subsection{Timing}
\label{subsec:tim}

\subsubsection{Persistent emission characteristics}
\label{subsec:tim_pers}

\begin{figure}
	\centering
	\includegraphics[scale=0.3, trim=0cm 0cm 0cm 0cm, clip=true]{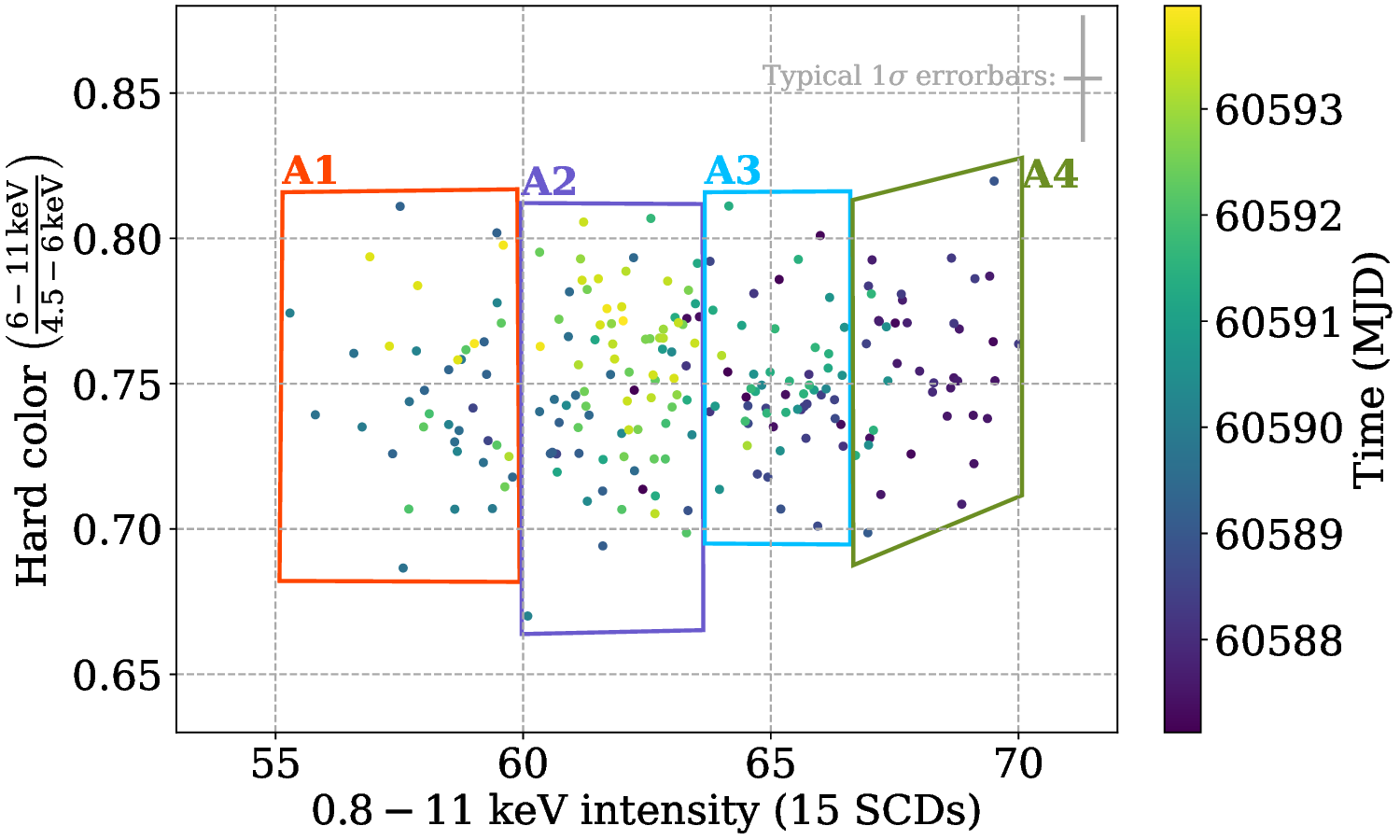}
    \caption{Hardness-intensity diagram of \aql over the entire observation (the burst and superburst durations have been filtered out). Each data point is 500~s. The intensity is defined as the count rate in $0.8-11$~keV and hard color is the ratio of count rates in $6-11$~keV and $4.5-6$~keV. The count rates shown are after background subtraction.}
    \label{fig:hidaql}
\end{figure}

\begin{figure}
	\centering
	\includegraphics[scale=0.3, trim=0cm 0cm 0cm 0cm, clip=true]{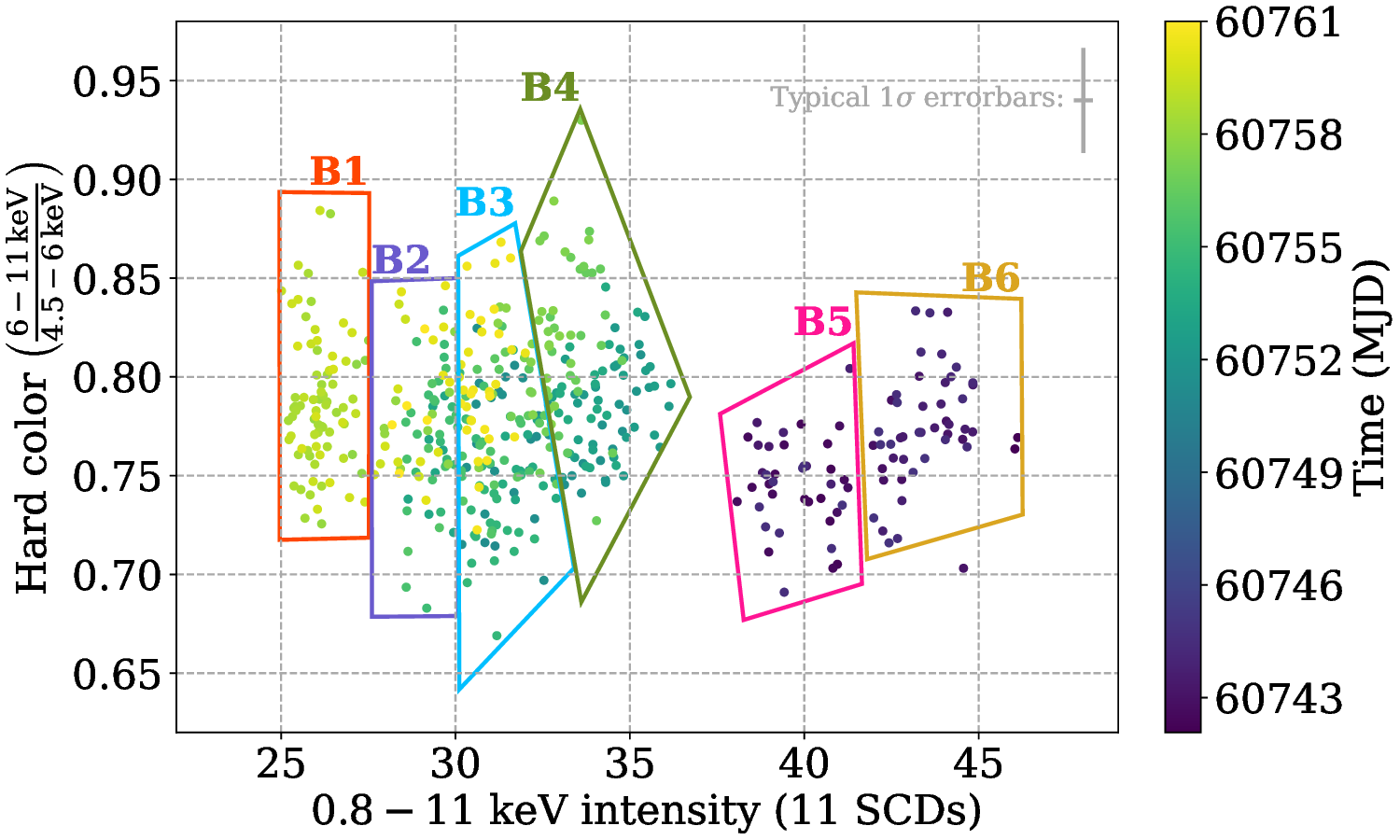}
    \caption{Same as Figure~\ref{fig:hidaql} but for \fouru}
    \label{fig:hid4u}
\end{figure}

Since the observation of both the sources are carried out during the outburst decay phase, we expect an overall decrease in the count rate with time. From the \textit{MAXI} light curves (Figure~\ref{fig:maxiaql4u}), the outburst decay lasted for $\sim 40$~d and $\sim 50$~d for \aql and \fouru respectively before returning to the pre-outburst levels. Over the \xsp observation span, the average background subtracted $0.8-11$~keV count rate decayed from 70~cps to 55~cps for \aql and 63~cps to 35~cps for \fouru.

First, to study the persistent emission from the sources, we removed the burst and superburst durations from the data, and constructed the CCD and HID as explained in Section~\ref{sec:obslis}. The HIDs of \aql and \fouru is shown in Figures~\ref{fig:hidaql} and \ref{fig:hid4u} respectively. The markers are color-coded with time to follow the evolution of the source throughout the observations. Based on the movement of the source in different regions of the HID, we defined 4 regions (A1 to A4) for \aql and 6 regions (B1 to B6) for \fouru, which are also marked in the figures. 

For \fouru, regions B5 and B6 correspond to the first observation segment when the source was brighter, and hence clearly separated from the other four regions on the intensity axis. B5 and B6 also have a lower spread in hardness as compared to B1 - B4. However, the average hardness remains comparable across the regions and do not show much variation. For \aql, the observation spanned a shorter duration, just beyond the peak of the outburst. This is evident from its HID (Figure~\ref{fig:hidaql}) which is essentially flat, implying no significant region-to-region hardness variations. For both the sources, due to the intensity modulation on the scale of days, the source keeps moving across the different regions, since the regions are majorly governed by the intensity rather than the hardness.

\subsubsection{Type-I X-ray bursts}
\label{subsec:tim_bur}

\begin{figure*}
\centering
\begin{subfigure}{0.33\textwidth}
\includegraphics[scale=0.29]{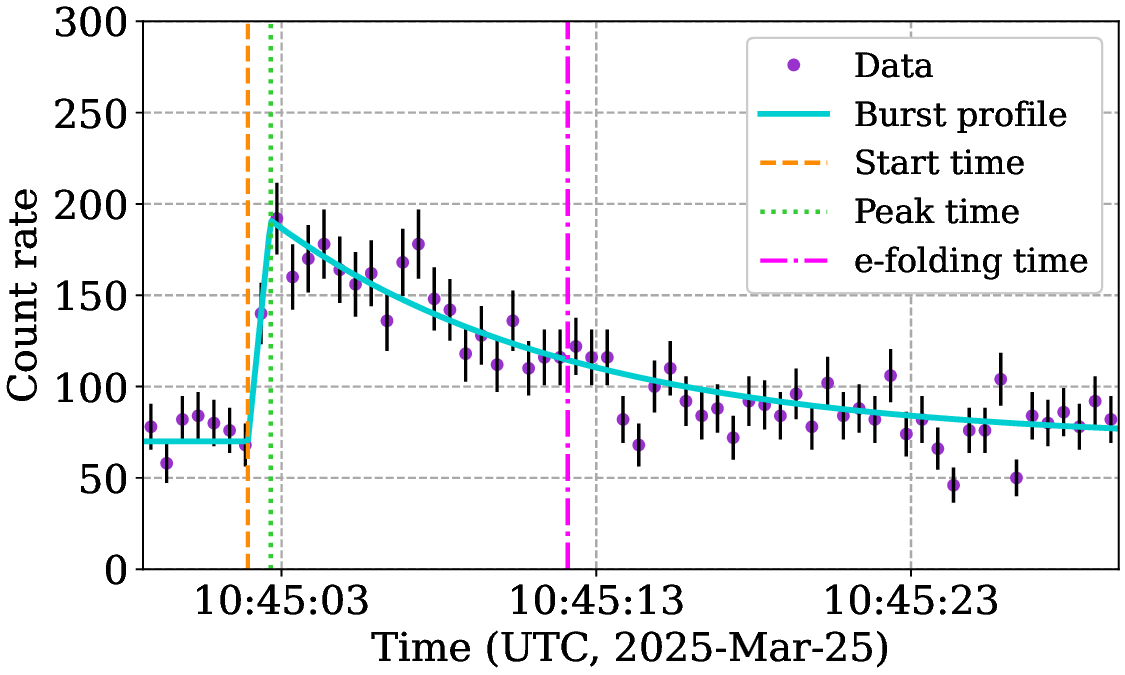}
\caption{}
\end{subfigure}\hfill
\begin{subfigure}{0.33\textwidth}
\includegraphics[scale=0.29]{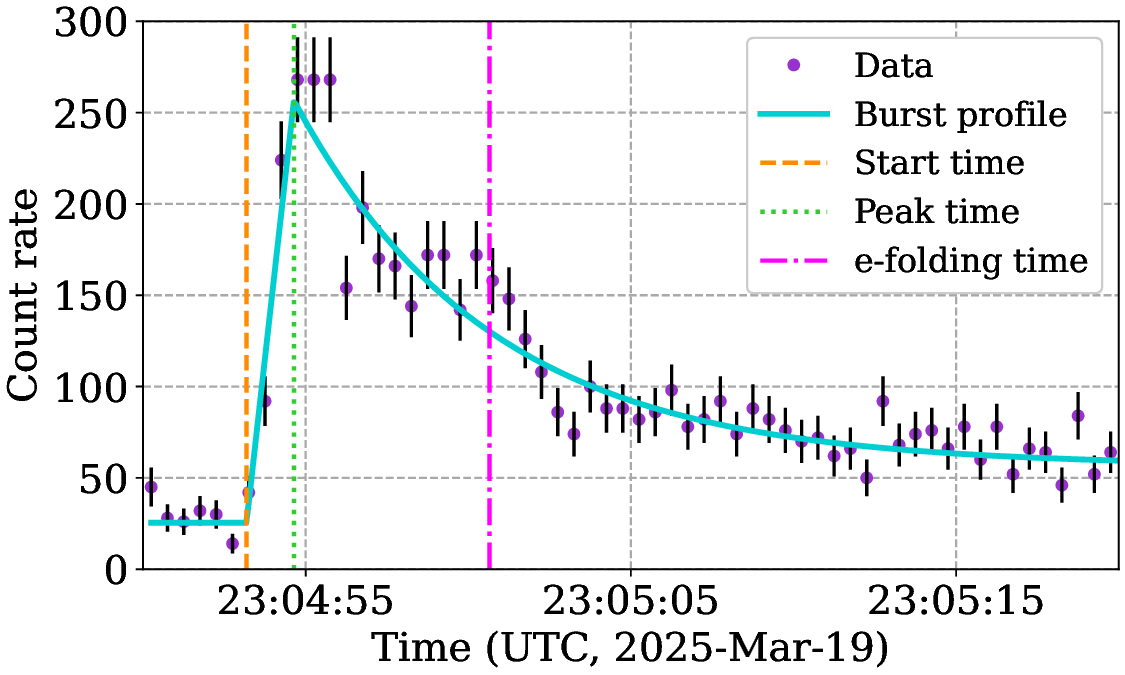}
\caption{}
\end{subfigure}\hfill
\begin{subfigure}{0.33\textwidth}
\includegraphics[scale=0.29]{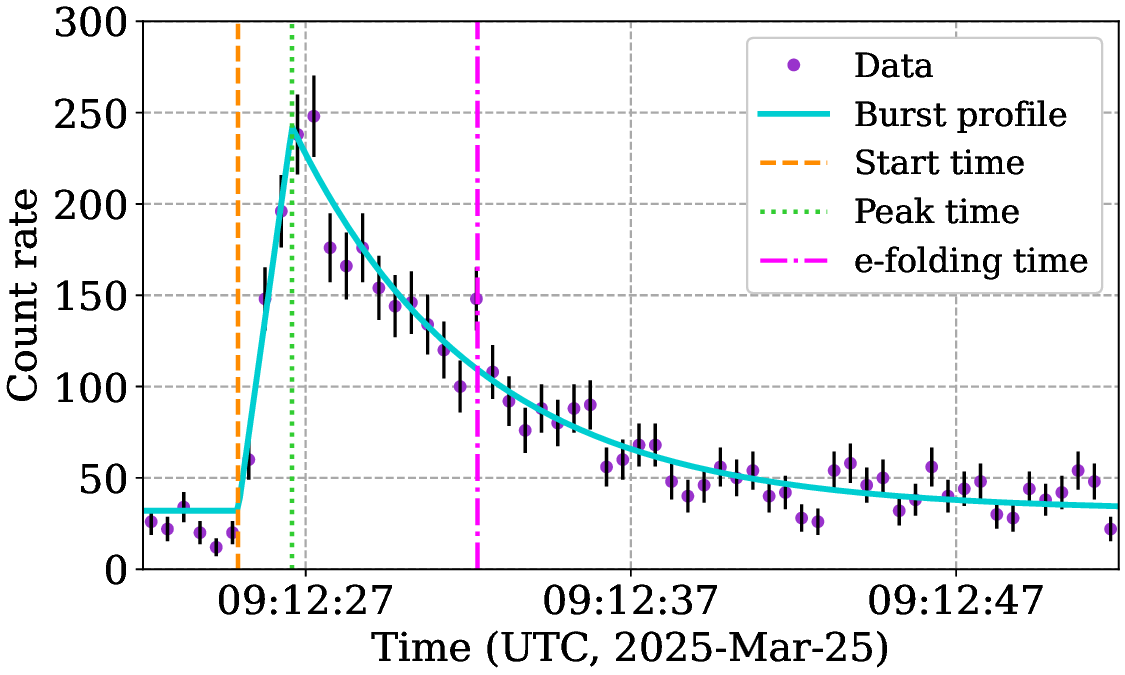}
\caption{}
\end{subfigure}  
\caption{Type-I X-ray bursts during \xsp observations of the two sources: (a) AqB1, (b) 4UB1, and (c) 4UB2. The bursts are plotted at 0.5~s binning and the count rates correspond to 15 detectors for \aql and 11 detectors for \fouru. The best-fit FRED profile is also plotted, with the start, peak, and e-folding times shown with vertical dashed, dotted, and dash-dotted lines respectively.}
\label{fig:burstfits}
\end{figure*} 

For all the three detected thermonuclear Type-I X-ray bursts, one from \aql (AqB1) and two from \fouru (4UB1 and 4UB2), we extracted the light curves at 0.5~second cadence and fitted them with a FRED (fast-rise-exponential decay) profile, as shown in Figure~\ref{fig:burstfits}. To account for the baseline persistent emission, we also included a constant offset. Notably, for 4UB1 which occurs $\sim$20~minutes before the superburst (see Section~\ref{subsec:tim_sb}), a single constant is unable to produce a good fit. This is because, as also evident from Figure~\ref{fig:burstfits}b, the pre-burst and post-burst persistent levels are significantly different. Hence, we included two constant factors to account for pre- and post-burst levels respectively. Moreover, for this burst, there appears to be hints of a second peak within $5-6$~s of the main peak. However, due to insufficient statistics, we do not attempt to model it further and go ahead with a simple FRED fit as for the other two bursts.

The bursts occur when the persistent fluxes are $1.7\times 10^{-8}$~erg~s$^{-1}$~cm$^{-2}$ (AqB1), $1.8\times 10^{-8}$~erg~s$^{-1}$~cm$^{-2}$ (4UB1), and $1.4\times 10^{-8}$~erg~s$^{-1}$~cm$^{-2}$ (4UB2). Assuming an Eddington luminosity $L_{\mathrm{Edd}}=2\times 10^{38}$~erg~s$^{-1}$ (\citealt{kuulkers2003}, consistent with a canonical 1.4~M$_{\odot}$ neutron star accreting hydrogen-rich material), these correspond to luminosites of $\sim 0.25$, $\sim 0.11$, and $\sim 0.08$~$L_{\mathrm{Edd}}$. The rise times range from $1-2$~s, with an e-folding time of 9.4~s for \aql and $\sim 6$~s for \fouru.

Various burst parameters such as burst start time (t$_{\mathrm{start}}$), rise duration (T$_{\mathrm{rise}}$), and decay time (T$_{\mathrm{decay}}$) are summarized in Table~\ref{tab:burst_params}. T$_{\mathrm{decay}}$ is computed as the time (from the burst peak time) taken for the burst flux to fall to 10\% of the peak burst flux\footnote{Flux above the background}. This can be calculated as $\tau\ln\left(\frac{1}{f}\right)$, where $\tau$ is the e-folding time and $f$ is the fraction of peak flux ($=0.1$ in this case). The pre-burst and post-burst count rates are also listed. 

\subsubsection{Superburst of \fouru}
\label{subsec:tim_sb}

\begin{figure}
	\centering
	\includegraphics[scale=0.35, trim=0cm 0cm 0cm 0cm, clip=true]{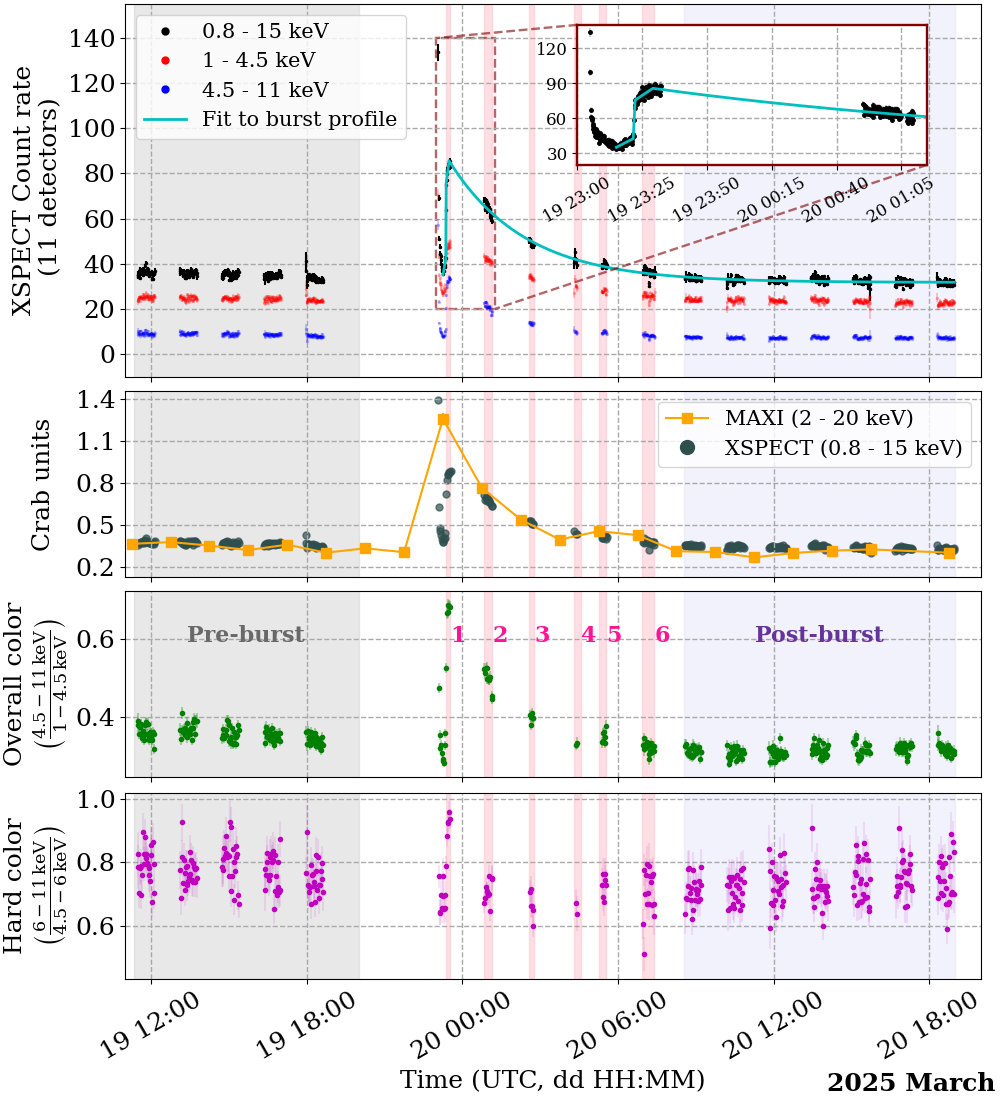}
    \caption{(Panel 1) Superburst lightcurve of \fouru in $1-4.5$~keV (red), $4.5-11$~keV (blue), and $0.8-15$~keV (black) energy bands, plotted with a bin size of 50~s. The fitted burst rise (piecewise linear) and decay (exponential) models are shown with solid cyan line. The inset shows a zoomed in view of the precursor burst and the first $\sim 1.5$~h of the superburst at 10~s binning. (Panel~2) MAXI ($2-20$~keV) observations are plotted in Crab units, with the \xsp $0.8-15$~keV light curve overlaid for comparison. The overall color (panel 3) defined as the ratio of count rates in $4.5-11$~keV and $1-4.5$~keV, and the hard color (panel 4) defined as the ratio of count rates in $6-11$~keV and $4.5-6$~keV are also shown. The pre-superburst, post-superburst, and superburst intervals $1-6$ are shaded with grey, lavender, and pink shades respectively. }
    \label{fig:superburst}
\end{figure}

The entire superburst (rise and decay) is covered by \xsp, for $10-15$~mins every orbit of XPoSat, with gaps of $\sim 60-70$~mins in between. This is because source observations are carried out only when the Sun is eclipsed by the Earth with respect to XPoSat \citep{saini2025}, and this gives a source observation duration ranging from $\sim 5-35$~mins every orbit (depending on the source location). Additionally, the detectors are switched off during SAA passage (and lead to a reduction in source view time if the SAA coincides with the eclipse duration), and two eclipse passes per day is reserved for data download, when the source of interest is not in the instrument FOV. All these constraints prevent in having a continuous coverage of the source.

The superburst begins in the orbit marked as `1' in Figure~\ref{fig:superburst}. The pre-burst $0.8-15$~keV count rate of the source was $\sim 35$~cps. Unfortunately, the two orbits preceding the superburst happened to be data dump orbits. Thus, what we define as `pre-burst' duration is $\sim 4.5$~h before the event.

In orbit~1, a thermonuclear (type-I) burst is first observed at $\sim$~23:05 UTC on 2025~March~19, reaching a peak count rate of $\sim 270$~cps (for 11 SCDs). This is also shown in Figure~\ref{fig:burstfits}b. This is a `precursor' burst to the superburst, and such bursts have been observed for several of the superbursts whose onsets have been captured. Like typical type-I bursts, it decayed exponentially, however it did not reach the pre-burst level (see Section~\ref{subsec:tim_bur}). Rather, it remained at an elevated level, decaying much slower than expected for $\sim 10$~mins, and then the source started brightening again. Within 7~mins of this (at $\sim$~23:22~UTC), the flux abruptly increased by a factor of 2, to $\sim 85$~cps. Subsequently, its decay to the pre-burst level occurred over the next six XPoSat orbits. Due to the gaps in the observing window, the upper limit of the time taken from the superburst peak to reach the pre-burst level is estimated to be $\sim 9$~h.

The \xsp light curve of the superburst of \fouru in different energy bands is shown in the top panel of Figure~\ref{fig:superburst}. The inset in the top panel of Figure~\ref{fig:superburst} shows the zoomed in light curve, which clearly shows the type-I burst decay, followed by the superburst onset, spaced by $\sim 20$~minutes. The second panel shows the simultaneous MAXI $2-20$~keV light curve (in Crab units) of the source, with the \xsp $0.8-15$~keV light curve (converted to Crab units) overlaid for comparison. The \xsp and MAXI count rates are consistent over the entire duration. 

The overall color, defined as ratio of counts in $4.5-11$~keV and $1-4.5$~keV is shown in the third panel, and the hard color (defined in Section~\ref{sec:obslis}) is shown in the bottom panel of Figure~\ref{fig:superburst}. The overall color follows the light curve, which suggests that at the peak of the superburst, the $4.5-11$~keV flux increases more than the $1-4.5$~keV implying a relative hardening in these bands. On the other hand, the hard color increases initially, coinciding with the superburst peak, but then dips slightly below the pre- and post-burst levels.

Unlike type-I bursts, where the fast rise can be well-fitted with a simple linear function, we observed a more complicated profile during the rising phase of the superburst. Since most superbursts have been observed with all-sky monitors which have a low duty cycle, the onsets have been captured in only a subset of them. In this case, we have modeled the rise of the superburst (following the decay of the precursor burst) using three piecewise linear functions, with a fast rise (1.15~cps~s$^{-1}$) over $\sim 27$~s, sandwiched between two relatively slower rises (0.02~cps~s$^{-1}$) of 6.5~mins each. This gives an estimate of the total rise time of $\sim 13.6$~mins for the superburst. The end-points of the piecewise functions were left as free parameters in our model, and the best fit peak time of the superburst was obtained as 23:28:37 UTC on 2025~March~19, which is $\sim 4$~mins before the orbit ended. This implies that the peak of the superburst was indeed observed by \xsp, with a peak count rate of $\sim 85$~cps. The superburst decay is modeled with an exponential decay function, which gives a best-fit e-folding time of 3.05~h. This model produces a good fit to the data and is shown in the top panel of Figure~\ref{fig:superburst}. The inferred T$_{\mathrm{decay}}$, as defined in Section~\ref{subsec:tim_bur}, is $\sim 7$~h. These parameters are summarized in Table~\ref{tab:burst_params}.

To see if different spectral components of the superburst behave differently, we fit the same model to lightcurves in the $1-4.5$~keV and $4.5-11$~keV bands. We find that compared to the superburst peak time in the full band ($0.8-15$~keV) mentioned above, the hard band emission peaks $2.6$~mins earlier and the soft band  $7.7$~mins later. The hard emission also decays faster (e-folding time 2.3~h) than the soft (3.6~h).

\subsubsection{Search for QPO}
\label{subsec:tim_qpo}

To check for quasi-periodic oscillations, we produced light curves with two different segment sizes and binsizes, to sample the different frequencies. To search for mHz QPOs, we produced 50~ms light curves in 512~s segments. This gives a frequency resolution of $\sim 1.9$~mHz, with a Nyquist frequency of 10~Hz. We also generated lightcurves with the finest possible binning of 1~ms, in 10~s segments. This covers the range of 0.1~Hz to 500~Hz, with 500~Hz being the highest possible frequency which can be probed by \xsp. For the persistent emission of both the sources, we produced day-wise power density spectra (PDS) in $0.8-3$~keV, $3-12$~keV, and $0.8-12$~keV, using the light curves summed across FOV. We also separately generated the PDS for the three type-I bursts and the superburst durations in these three energy bands. No QPOs are detected in the PDS during either the persistent or burst durations.

\subsection{Spectral}
\label{subsec:spec}

\subsubsection{Persistent emission}
\label{subsec:spec_pers}

\begin{figure*}
\centering
\begin{subfigure}{0.33\textwidth}
\includegraphics[scale=0.36]{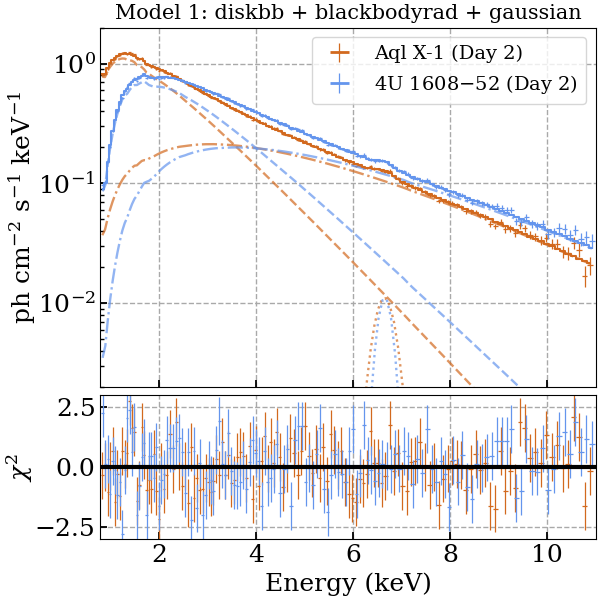}
\caption{}
\end{subfigure}\hfill
\begin{subfigure}{0.33\textwidth}
\includegraphics[scale=0.36]{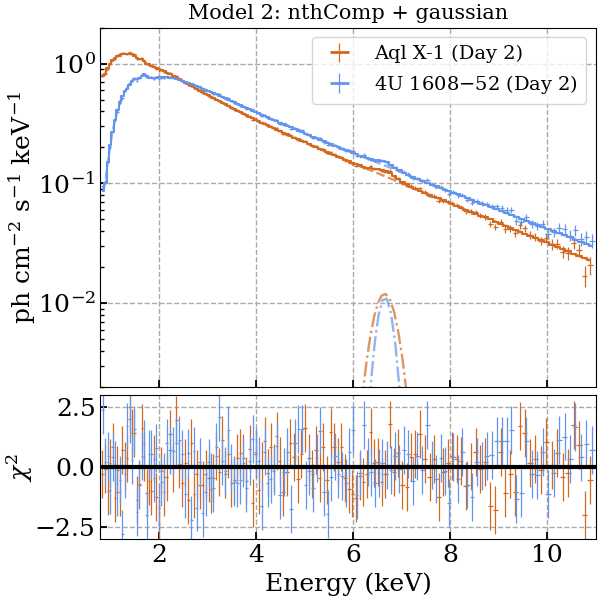}
\caption{}
\end{subfigure}\hfill
\begin{subfigure}{0.33\textwidth}
\includegraphics[scale=0.36]{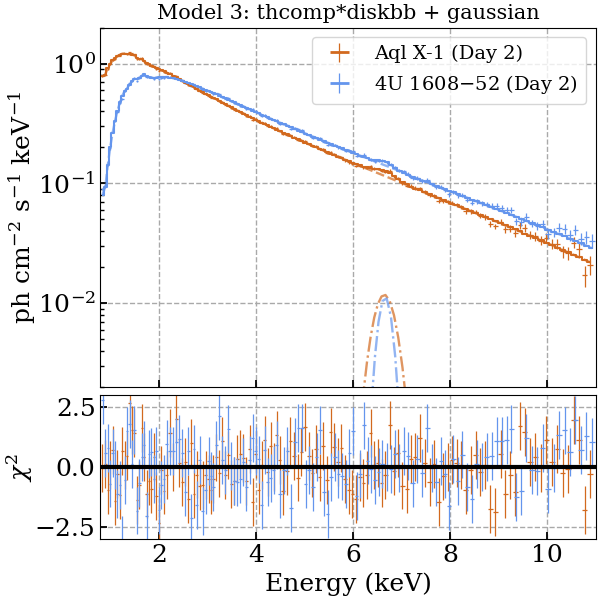}
\caption{}
\end{subfigure}  
\caption{Representative spectral fits for day~2 observations of both \aql \fouru using (a) model~1, (b) model~2, and  (c) model~3. The top panel is the unfolded spectra and the bottom panel show the residuals. Individual components of the models are also plotted. For clarity, only $2^{\circ}\times 2^{\circ}$ spectra are shown.}
\label{fig:specfit}
\end{figure*} 

As evident from the HID of \aql (Figure~\ref{fig:hidaql}), there is not much variation in the hardness over the observation. Moreover, the intensity change from A1 to A4 is only about 1~cps~per~SCD. Both these point to the fact that the spectral state of the source remained nearly constant and we do not expect significant variations in the spectra of the four branches. For \fouru, the two observation segments were separated  by a week, so the intensity decay is evident, along with a spread in hardness. However, as we do not observe much variation in the hardness for both sources, we extracted day-wise spectra (bottom panels of Figures~\ref{fig:fullaql} and \ref{fig:full4u} show the corresponding time intervals) to study the evolution of persistent emission across the observation period. We attempted fitting the spectra using several continuum models. We present three of the models here, which provided the best fits to the spectra of both the sources (with respect to goodness-of-fit) with physically plausible values of parameters:
\begin{itemize}
\item \texttt{diskbb + bbodyrad} (model~1): A combination of multi-color blackbody component attributed to the accretion disk \citep{mitsuda1984}, along with a single temperature blackbody component attributed to the emission from the neutron star surface. Each component has two parameters (inner disk temperature $kT_{in}$ and blackbody temperature $kT_{BB}$, along with respective normalizations). The normalization factor of the two components can be used to estimate the size of the emission regions. The color-corrected inner radius of the accretion disk $R_{in}$ (in km) can be estimated as $f^2R_{app}$, where $f$ is the spectral hardening factor (\citealt{davis2019, agrawal2025}), and the apparent radius $R_{app}$ is given by
\begin{equation}
R_{app} = \sqrt{\frac{N_{disk}d_{10}^2}{\cos\theta}}.
\end{equation}
Here, $N_{disk}$ is the disk normalization, $\theta$ is the orbital inclination of the system and $d_{10}$ is the distance to the source in units of 10~kpc. The value of $f$ is found to be in the range $1.5-1.6$. The blackbody radius $R_{BB}$ (in km) is calculated as
\begin{equation}
R_{BB} = \sqrt{N_{BB}d_{10}^2}
\end{equation}
where $N_{BB}$ is the normalization of the \texttt{bbdodyrad} component.

\item \texttt{nThcomp} (model~2): A thermal Comptonization model, which results from the upscattering of seed photons (either from the NS or disk) by the hot plasma in the corona (\citealt{zdziarski1996, zycki1999}). The continuum shape is described by the photon index $\Gamma$ and electron temperature $kT_e$. The seed photons are parametrized by the seed photon temperature $kT_s$. Note that we have assumed the seed photons to originate from the NS (\texttt{inp\_type} = 0). For both the sources, this single continuum model provided a good fit to the spectra of all the days ($\chi^2_{\mathrm{red}}$ in the range $0.8-1.2$). Under the assumption of a spherically symmetric, uniform density corona, the optical depth $\tau$ of the Comptonization region is related to $\Gamma$ and $kT_e$ as \citep{zdziarski1996}:
\begin{equation}
\Gamma = -\frac{1}{2} + \left[\frac{9}{4} + \frac{1}{\frac{kT_e}{m_ec^2}\left(1+\frac{\tau}{3}\right)\tau} \right]^{1/2}
\end{equation}

The Compton y-parameter, which represents the degree of modification of the seed photon spectra due to Comptonization, is defined in the optically thick limit as
\begin{equation}
y=\frac{4kT_e\tau^2}{m_ec^2}.
\end{equation} 

\item \texttt{thcomp*diskbb} (model~3): \texttt{thcomp} is a convolution model which, when applied to a seed photon distribution, describes its thermal Comptonization spectrum. In this case, we have assumed the disk to provide the seed photons which are then Comptonized by thermal electrons. The parameters $\Gamma$ and $kT_{e}$ are similar to those in \texttt{nThcomp} (model~2). In addition, the covering fraction ($cov\_frac$), a number between 0 and 1, controls the fraction of seed photons which are Comptonized. Two $diskbb$ parameters are also present, as described in model~1. In all our fits, $cov\_frac$ was between $0.9-1$, but its uncertainty could not be constrained, so we froze it to 1. All other parameters were allowed to vary freely during the fits.

\end{itemize}

For each of these three models, a Gaussian component at $\sim 6.7$~keV was required for both the sources, and its inclusion improved the fits significantly (F-test probability of $< 10^{-6}$ for chance improvement due to added component). The continuum and line components were modified for interstellar absorption along the line of sight using the Tuebingen-Boulder ISM absorption model, with abundances from \cite{wilms2000}. As mentioned in Section~\ref{sec:obslis}, a constant multiplicative factor to account for the cross-calibration between the two FOVs, and two \texttt{edge} components corresponding to absorption edges of Al and Si were also included. The final three models in \textit{XSPEC} are:

\begin{enumerate}[label=Model~\arabic*:,  leftmargin=\widthof{[Model~1: ]}+\labelsep ] 
\item \texttt{TBabs $\times$ constant $\times$ edge $\times$ edge $\times$ (diskbb + bbodyrad + gaussian)}
\item \texttt{TBabs $\times$ constant $\times$ edge $\times$ edge $\times$ (nthComp + gaussian)} 
\item \texttt{TBabs $\times$ constant $\times$ edge $\times$ edge $\times$ (thcomp*diskbb + gaussian)}
\end{enumerate}

These three models give good fits to the daywise spectra of both the sources ($\chi^2_{red}$ in the range $0.84-1.18$ for \aql and $0.87-1.22$ for \fouru). Representive spectral fits for \aql and \fouru (respective day~2 of both sources) using all the three models are shown in Figure~\ref{fig:specfit}. Models 2 and 3 are both Comptonization models. Here, we prefer \texttt{thcomp} as it agrees better with actual Monte Carlo spectra from Comptonization, as compared to \texttt{nThcomp} \citep{zdziarski2020}. Moreover, the seed photon spectrum can be explicitly specified here, along with a covering fraction (even though it could not be constrained in our fits). Hence we go ahead with models~1 and 3. The best-fit parameters using these two models, as well as some derived parameters, are presented in Tables~\ref{tab:model1} and \ref{tab:model3}. The component-wise unabsorbed fluxes were calculated using $cflux$ in the $0.8-15$~keV energy range, which are also reported in the tables. All quoted uncertainties correspond to 1$\sigma$ (68\% c.l) errors.  

\begin{landscape}

\begin{table}
\centering
\begin{threeparttable}
\caption{Best-fit spectral parameters using model~1 (\texttt{diskbb~+~bbodyrad}) for both the sources, along with component-wise fluxes and total luminosity. The pre-superburst and post-superburst parameters are also reported (See Section~\ref{subsec:spec_sb}).}
\label{tab:model1}
\bgroup
\def\arraystretch{1.4}
\setlength{\tabcolsep}{2.5pt}
\begin{tabular}{cccccccccccccccc}
\hline
\multirow{2}{*}{Time} & \multirow{2}{*}{\begin{tabular}[c]{@{}c@{}}$n_{H}$\tnote{*} \end{tabular}} & \multicolumn{3}{c}{diskbb}      & \multicolumn{2}{c}{blackbody}   & \multicolumn{3}{c}{gaussian}                                                      & \multirow{2}{*}{$\chi^2_{red}$ ($\chi^2/dof$)} & \multirow{2}{*}{\begin{tabular}[c]{@{}c@{}}$F_{disk}$\tnote{\textdaggerdbl}\end{tabular}} & \multirow{2}{*}{\begin{tabular}[c]{@{}c@{}}$F_{BB}$\tnote{\textdaggerdbl}\end{tabular}} & \multirow{2}{*}{\begin{tabular}[c]{@{}c@{}}$F_{gauss}$\tnote{\textdaggerdbl}\end{tabular}} & \multirow{2}{*}{\begin{tabular}[c]{@{}c@{}}$L_{tot}$\tnote{\S}\end{tabular}} & \multirow{2}{*}{\begin{tabular}[c]{@{}c@{}}$L_{tot}/L_{Edd}$\tnote{\S}\end{tabular}} \\
  &  & $kT_{in}$ (keV) & $R_{app}$ (km) & $R_{in}$ (km) & $kT_{BB}$ (keV) & $R_{BB}$ (km) & $E_{g}$ (keV) & $\sigma_{g}$ (keV) & $N_{g}$\tnote{\textdagger}  &  &  &  &  &  & \\ \hline
                      \multicolumn{16}{c}{Aql X$-$1}                                                                                                                                                                                                                                                                                                                                                                                                                                                                                                                                                                                                                                                    \\ \hline
Day 1 (Oct 4) & $0.39^{+0.01}_{-0.0}$ & $0.91^{+0.02}_{-0.02}$ & $15.7 \pm 2.9$ & $42.0 \pm 7.7$ & $1.65^{+0.01}_{-0.01}$ & $5.8 \pm 1.1$ & $6.61^{+0.04}_{-0.04}$ & $0.16^{+0.04}_{-0.05}$ & $0.38^{+0.08}_{-0.08}$ & $1.07\ (279.5/259)$ & $9.26 \pm 0.02$ & $10.50 \pm 0.03$ & $0.040 \pm 0.006$ & $5.92 \pm 2.13$ & $0.30 \pm 0.11$ \\ 
Day 2 (Oct 5) & $0.37^{+0.01}_{-0.01}$ & $0.92^{+0.02}_{-0.02}$ & $15.2 \pm 2.8$ & $40.4 \pm 7.4$ & $1.65^{+0.01}_{-0.01}$ & $5.6 \pm 1.0$ & $6.65^{+0.03}_{-0.03}$ & $0.2^{+0.03}_{-0.03}$ & $0.56^{+0.09}_{-0.08}$ & $0.99\ (257.4/258)$ & $9.10 \pm 0.02$ & $9.90 \pm 0.03$ & $0.060 \pm 0.007$ & $5.70 \pm 2.05$ & $0.29 \pm 0.10$ \\
Day 3 (Oct 6) & $0.37^{+0.01}_{-0.01}$ & $0.89^{+0.01}_{-0.01}$ & $16.0 \pm 2.9$ & $41.5 \pm 7.6$ & $1.61^{+0.01}_{-0.01}$ & $5.4 \pm 1.0$ & $6.69^{+0.03}_{-0.03}$ & $0.12^{+0.05}_{-0.04}$ & $0.30^{+0.07}_{-0.06}$ & $1.06\ (274.3/256)$ & $8.50 \pm 0.02$ & $8.38 \pm 0.03$ & $0.032 \pm 0.005$ & $5.06 \pm 1.82$ & $0.25 \pm 0.09$ \\ 
Day 4 (Oct 7) & $0.37^{+0.01}_{-0.01}$ & $0.93^{+0.02}_{-0.02}$ & $15.0 \pm 2.7$ & $39.5 \pm 7.2$ & $1.65^{+0.02}_{-0.02}$ & $5.4 \pm 1.0$ & $6.6^{+0.04}_{-0.04}$ & $0.14^{+0.04}_{-0.03}$ & $0.33^{+0.07}_{-0.07}$ & $0.90\ (232.4/255)$ & $9.00 \pm 0.02$ & $9.12 \pm 0.03$ & $0.034 \pm 0.006$ & $5.43 \pm 1.96$ & $0.27 \pm 0.10$ \\ 
Day 5 (Oct 8) & $0.37^{+0.01}_{-0.01}$ & $0.92^{+0.02}_{-0.02}$ & $15.1 \pm 2.8$ & $39.9 \pm 7.3$ & $1.64^{+0.02}_{-0.02}$ & $5.6 \pm 1.0$ & $6.56^{+0.05}_{-0.05}$ & $0.24^{+0.05}_{-0.04}$ & $0.51^{+0.10}_{-0.10}$ & $1.15\ (297.1/257)$ & $8.84 \pm 0.02$ & $9.65 \pm 0.03$ & $0.054 \pm 0.008$ & $5.55 \pm 2.00$ & $0.28 \pm 0.10$ \\
Day 6 (Oct 9) & $0.38^{+0.01}_{-0.01}$ & $0.9^{+0.01}_{-0.01}$ & $15.6 \pm 2.8$ & $40.9 \pm 7.5$ & $1.64^{+0.01}_{-0.01}$ & $5.5 \pm 1.0$ & $6.59^{+0.03}_{-0.03}$ & $0.18^{+0.03}_{-0.03}$ & $0.51^{+0.08}_{-0.08}$ & $0.98\ (252.9/256)$ & $8.67 \pm 0.02$ & $8.96 \pm 0.03$ & $0.054 \pm 0.006$ & $5.29 \pm 1.91$ & $0.26 \pm 0.10$ \\ 
Day 7 (Oct 10) & $0.38^{+0.01}_{-0.0}$ & $0.92^{+0.01}_{-0.02}$ & $14.8 \pm 2.7$ & $38.9 \pm 7.1$ & $1.66^{+0.01}_{-0.02}$ & $5.4 \pm 1.0$ & $6.5^{+0.04}_{-0.04}$ & $0.17^{+0.04}_{-0.03}$ & $0.42^{+0.08}_{-0.08}$ & $1.17\ (302.0/256)$ & $8.61 \pm 0.02$ & $9.22 \pm 0.03$ & $0.044 \pm 0.006$ & $5.35 \pm 1.93$ & $0.27 \pm 0.10$ \\  \hline
\multicolumn{16}{c}{4U $1608-52$}                                                                                                                                                                                                                                                                                                                                                                                                                                                                                                                                                                                                                                                 \\ \hline
Pre-superburst & $1.06^{+0.02}_{-0.02}$ &  $0.98^{+0.04}_{-0.03}$  &  $8.7 \pm 1.0$ & $19.3 \pm 2.2$ & $1.73^{+0.03}_{-0.03}$  &  $3.0 \pm 0.3$  &  $6.75^{+0.07}_{-0.05}$  &  $0.12^{+0.05}_{-0.08}$  & $0.35^{+0.10}_{-0.11}$ & $1.05\ (253.6/241)$ & $9.51 \pm 0.03$ & $8.46 \pm 0.04$ & $0.038 \pm 0.008$ & $2.21 \pm 0.41$ & $0.11 \pm 0.02$ \\
Post-superburst & $1.06^{+0.02}_{-0.02}$ &  $1.00^{+0.03}_{-0.02}$  &  $8.6 \pm 0.9$ & $18.7 \pm 2.0$ & $1.71^{+0.04}_{-0.03}$  &  $2.7 \pm 0.3$  &  $6.70^{+0.04}_{-0.04}$  &  $0.13^{+0.05}_{-0.05}$  & $0.32^{+0.08}_{-0.08}$ & $1.01\ (248.0/246)$ & $9.94 \pm 0.03$ & $6.28 \pm 0.03$ & $0.028 \pm 0.006$ & $1.99 \pm 0.37$ & $0.10 \pm 0.02$ \\
Day 1 (Mar 8) & $1.08^{+0.03}_{-0.01}$ & $0.97^{+0.02}_{-0.04}$ & $9.8 \pm 1.1$ & $22.8 \pm 2.6$ & $1.69^{+0.02}_{-0.03}$ & $3.6 \pm 0.4$ & $6.74^{+0.06}_{-0.06}$ & $0.2^{+0.07}_{-0.05}$ & $0.47^{+0.12}_{-0.12}$ & $1.04\ (259.3/247)$ & $11.49 \pm 0.04$ & $10.37 \pm 0.04$ & $0.051 \pm 0.010$ & $2.68 \pm 0.50$ & $0.13 \pm 0.03$ \\
Day 2 (Mar 9) & $1.04^{+0.02}_{-0.01}$ & $1.03^{+0.03}_{-0.04}$ & $8.8 \pm 1.0$ & $20.6 \pm 2.4$ & $1.77^{+0.03}_{-0.03}$ & $3.4 \pm 0.4$ & $6.66^{+0.05}_{-0.05}$ & $0.14^{+0.08}_{-0.06}$ & $0.41^{+0.13}_{-0.11}$ & $1.09\ (273.8/249)$ & $11.87 \pm 0.04$ & $11.34 \pm 0.05$ & $0.044 \pm 0.009$ & $2.85 \pm 0.53$ & $0.14 \pm 0.03$ \\
Day 3 (Mar 10) & $1.03^{+0.04}_{-0.0}$ & $1.01^{+0.03}_{-0.03}$ & $8.9 \pm 1.0$ & $20.7 \pm 2.3$ & $1.75^{+0.03}_{-0.03}$ & $3.5 \pm 0.4$ & $6.7^{+0.03}_{-0.04}$ & $0.11^{+0.06}_{-0.04}$ & $0.43^{+0.13}_{-0.12}$ & $1.21\ (307.3/251)$ & $11.64 \pm 0.03$ & $11.13 \pm 0.04$ & $0.046 \pm 0.008$ & $2.80 \pm 0.52$ & $0.14 \pm 0.03$ \\
Day 4 (Mar 17) & $1.03^{+0.02}_{-0.08}$ & $1.01^{+0.03}_{-0.02}$ & $8.2 \pm 0.9$ & $18.2 \pm 1.9$ & $1.74^{+0.03}_{-0.03}$ & $2.8 \pm 0.3$ & $6.62^{+0.05}_{-0.05}$ & $0.2^{+0.05}_{-0.04}$ & $0.43^{+0.09}_{-0.08}$ & $1.15\ (288.6/248)$ & $9.60 \pm 0.03$ & $7.23 \pm 0.03$ & $0.045 \pm 0.007$ & $2.07 \pm 0.39$ & $0.10 \pm 0.02$ \\
Day 5 (Mar 18) & $1.07^{+0.02}_{-0.02}$ & $0.95^{+0.03}_{-0.01}$ & $9.0 \pm 1.0$ & $20.3 \pm 2.2$ & $1.7^{+0.02}_{-0.02}$ & $3.3 \pm 0.3$ & $6.66^{+0.05}_{-0.04}$ & $0.17^{+0.05}_{-0.05}$ & $0.41^{+0.10}_{-0.09}$ & $1.16\ (290.6/249)$ & $9.21 \pm 0.03$ & $9.45 \pm 0.03$ & $0.044 \pm 0.007$ & $2.29 \pm 0.43$ & $0.11 \pm 0.02$ \\
Day 6 (Mar 19) & $1.04^{+0.03}_{-0.01}$ & $1.0^{+0.03}_{-0.03}$ & $8.4 \pm 0.9$ & $18.6 \pm 2.0$ & $1.74^{+0.03}_{-0.03}$ & $2.9 \pm 0.3$ & $6.69^{+0.02}_{-0.04}$ & $0.2^{+0.05}_{-0.04}$ & $0.46^{+0.10}_{-0.10}$ & $1.16\ (291.7/250)$ & $9.68 \pm 0.03$ & $7.46 \pm 0.03$ & $0.049 \pm 0.007$ & $2.11 \pm 0.40$ & $0.11 \pm 0.02$ \\
Day 7 (Mar 20) & $1.06^{+0.03}_{-0.0}$ & $0.97^{+0.03}_{-0.02}$ & $9.0 \pm 1.0$ & $19.6 \pm 2.1$ & $1.69^{+0.03}_{-0.03}$ & $2.8 \pm 0.3$ & $6.76^{+0.01}_{-0.02}$ & $0.15$\tnote{$\|$} & $0.26^{+0.06}_{-0.06}$ & $0.98\ (240.4/244)$ & $9.73 \pm 0.03$ & $6.32 \pm 0.03$ & $0.021 \pm 0.004$ & $1.97 \pm 0.37$ & $0.10 \pm 0.02$ \\
Day 8 (Mar 21) & $1.04^{+0.02}_{-0.01}$ & $0.98^{+0.02}_{-0.03}$ & $8.5 \pm 0.9$ & $18.5 \pm 1.9$ & $1.73^{+0.02}_{-0.02}$ & $2.7 \pm 0.3$ & $6.69^{+0.03}_{-0.03}$ & $0.1^{+0.03}_{-0.03}$ & $0.28^{+0.06}_{-0.06}$ & $1.03\ (258.8/249)$ & $8.94 \pm 0.03$ & $6.76 \pm 0.03$ & $0.030 \pm 0.005$ & $1.93 \pm 0.36$ & $0.10 \pm 0.02$ \\
Day 9 (Mar 22) & $1.1^{+0.02}_{-0.02}$ & $0.92^{+0.02}_{-0.02}$ & $9.6 \pm 1.0$ & $21.5 \pm 2.3$ & $1.71^{+0.02}_{-0.02}$ & $3.2 \pm 0.3$ & $6.72^{+0.06}_{-0.06}$ & $0.15^{+0.05}_{-0.06}$ & $0.25^{+0.08}_{-0.08}$ & $1.18\ (297.2/250)$ & $8.82 \pm 0.03$ & $9.29 \pm 0.03$ & $0.027 \pm 0.007$ & $2.22 \pm 0.42$ & $0.11 \pm 0.02$ \\
Day 10 (Mar 23) & $1.06^{+0.02}_{-0.02}$ & $0.95^{+0.03}_{-0.02}$ & $8.7 \pm 0.9$ & $19.2 \pm 2.0$ & $1.76^{+0.02}_{-0.02}$ & $2.9 \pm 0.3$ & $6.67^{+0.05}_{-0.05}$ & $0.14^{+0.05}_{-0.05}$ & $0.26^{+0.07}_{-0.07}$ & $0.98\ (248.7/253)$ & $8.53 \pm 0.02$ & $8.45 \pm 0.03$ & $0.028 \pm 0.006$ & $2.08 \pm 0.39$ & $0.10 \pm 0.02$ \\
Day 11 (Mar 24) & $1.08^{+0.03}_{-0.01}$ & $0.91^{+0.02}_{-0.02}$ & $9.6 \pm 1.0$ & $20.4 \pm 2.1$ & $1.68^{+0.03}_{-0.02}$ & $2.6 \pm 0.3$ & $6.65^{+0.04}_{-0.04}$ & $0.13^{+0.04}_{-0.04}$ & $0.26^{+0.06}_{-0.06}$ & $0.86\ (215.3/247)$ & $8.22 \pm 0.02$ & $5.56 \pm 0.02$ & $0.027 \pm 0.005$ & $1.69 \pm 0.32$ & $0.08 \pm 0.02$ \\
Day 12 (Mar 25) & $1.03^{+0.02}_{-0.02}$ & $0.97^{+0.03}_{-0.02}$ & $8.3 \pm 0.9$ & $17.5 \pm 1.9$ & $1.73^{+0.03}_{-0.03}$ & $2.5 \pm 0.3$ & $6.71^{+0.05}_{-0.05}$ & $0.23^{+0.07}_{-0.07}$ & $0.37^{+0.10}_{-0.09}$ & $1.11\ (277.4/247)$ & $8.28 \pm 0.02$ & $5.65 \pm 0.03$ & $0.040 \pm 0.006$ & $1.71 \pm 0.32$ & $0.09 \pm 0.02$ \\
Day 13 (Mar 26) & $1.04^{+0.05}_{-0.01}$ & $0.96^{+0.03}_{-0.02}$ & $8.4 \pm 0.9$ & $18.5 \pm 2.0$ & $1.74^{+0.02}_{-0.02}$ & $3.0 \pm 0.3$ & $6.63^{+0.05}_{-0.05}$ & $0.14^{+0.06}_{-0.05}$ & $0.30^{+0.10}_{-0.11}$ & $1.03\ (259.9/250)$ & $8.51 \pm 0.02$ & $7.73 \pm 0.03$ & $0.034 \pm 0.006$ & $1.99 \pm 0.37$ & $0.10 \pm 0.02$ \\ \hline
\end{tabular}%
\egroup
\begin{tablenotes} 
 			\item[*] $\times 10^{22}$ cm$^{-2}$
            \item[\textdagger] $\times 10^{-2}$ ph~cm$^{-2}$~s$^{-1}$
            \item[\textdaggerdbl] $\times 10^{-9}$~erg~s$^{-1}$~cm$^{-2}$, unabsorbed, $0.8 - 15$ keV
            \item[\S]  $\times 10^{37}$~erg~s$^{-1}$  
            \item[$\|$] Could not be constrained, so frozen to average of other observations     
\end{tablenotes}
\end{threeparttable}
\end{table}

\end{landscape}

\begin{landscape}

\begin{table}
\centering
\begin{threeparttable}
\caption{Same as Table~\ref{tab:model1} but for model~3 (\texttt{thcomp~$\times$~diskbb}).}
\label{tab:model3}
\bgroup
\def\arraystretch{1.4}
\setlength{\tabcolsep}{2.5pt}
\begin{tabular}{ccccccccccccccccc}
\hline
\multirow{2}{*}{Time} & \multirow{2}{*}{\begin{tabular}[c]{@{}c@{}}$n_{H}$\tnote{*}\end{tabular}} & \multicolumn{2}{c}{thcomp}                                & \multicolumn{3}{c}{diskbb}      & \multicolumn{3}{c}{gaussian}                                                      & \multirow{2}{*}{$\chi^2_{red}$} & \multirow{2}{*}{$\tau$} & \multirow{2}{*}{$y$} & \multirow{2}{*}{\begin{tabular}[c]{@{}c@{}}$F_{th*disk}$\tnote{\textdaggerdbl}\end{tabular}} & \multirow{2}{*}{\begin{tabular}[c]{@{}c@{}}$F_{gauss}$\tnote{\textdaggerdbl}\end{tabular}} & \multirow{2}{*}{\begin{tabular}[c]{@{}c@{}}$L_{tot}$\tnote{\S}\end{tabular}} & \multirow{2}{*}{\begin{tabular}[c]{@{}c@{}}$L_{tot}/L_{Edd}$\tnote{\S}\end{tabular}} \\
                      &                                                                                           & \multicolumn{1}{c}{$\Gamma$} & \multicolumn{1}{c}{$kT_e$ (keV)} & $kT_{in}$ (keV) & $R_{app}$ (km) & $R_{in}$ (km) & $E_{g}$ (keV) & $\sigma_{g}$ (keV) & $N_{g}$\tnote{\textdagger} &                                 &                         &                      &                                                                                                &                                                                                          &                                                                                        \\ \hline
                      \multicolumn{17}{c}{Aql X$-$1}                                                                                                                                                                                                                                                                                                                                                                                                                                                                                                                                                                                                                                              \\ \hline
Day 1 (Oct 4) & $0.34^{+0.01}_{-0.0}$ & $1.82^{+0.01}_{-0.0}$ & $2.3^{+0.04}_{-0.03}$ & $0.54^{+0.01}_{-0.04}$ & $45.1 \pm 8.8$ & $120.2 \pm 23.5$ & $6.61^{+0.04}_{-0.04}$ & $0.2^{+0.05}_{-0.04}$ & $0.48^{+0.08}_{-0.08}$ & $1.05\ (273.9/259)$ & $13.2 \pm 0.2$ & $3.1 \pm 0.1$ & $19.25 \pm 0.02$ & $0.051 \pm 0.007$ & $5.78 \pm 2.08$ & $0.29 \pm 0.10$ \\
Day 2 (Oct 5) & $0.32^{+0.01}_{-0.01}$ & $1.84^{+0.01}_{-0.01}$ & $2.29^{+0.04}_{-0.03}$ & $0.56^{+0.02}_{-0.02}$ & $41.8 \pm 7.9$ & $110.5 \pm 20.9$ & $6.65^{+0.03}_{-0.02}$ & $0.22^{+0.03}_{-0.03}$ & $0.66^{+0.09}_{-0.08}$ & $0.96\ (249.3/258)$ & $13.0 \pm 0.3$ & $3.0 \pm 0.1$ & $18.52 \pm 0.02$ & $0.071 \pm 0.007$ & $5.56 \pm 2.00$ & $0.28 \pm 0.10$ \\
Day 3 (Oct 6) & $0.32^{+0.01}_{-0.0}$ & $1.9^{+0.0}_{-0.01}$ & $2.27^{+0.03}_{-0.01}$ & $0.54^{+0.0}_{-0.03}$ & $42.1 \pm 7.6$ & $109.0 \pm 19.7$ & $6.68^{+0.04}_{-0.04}$ & $0.19^{+0.05}_{-0.05}$ & $0.42^{+0.01}_{-0.08}$ & $1.05\ (270.1/256)$ & $12.4 \pm 0.2$ & $2.7 \pm 0.1$ & $16.41 \pm 0.02$ & $0.044 \pm 0.006$ & $4.92 \pm 1.77$ & $0.25 \pm 0.09$ \\
Day 4 (Oct 7) & $0.32^{+0.01}_{-0.01}$ & $1.87^{+0.01}_{-0.01}$ & $2.31^{+0.04}_{-0.02}$ & $0.56^{+0.01}_{-0.02}$ & $40.9 \pm 8.1$ & $107.2 \pm 21.1$ & $6.6^{+0.04}_{-0.04}$ & $0.15^{+0.02}_{-0.04}$ & $0.38^{+0.07}_{-0.04}$ & $0.86\ (222.0/255)$ & $12.6 \pm 0.2$ & $2.9 \pm 0.1$ & $17.63 \pm 0.02$ & $0.040 \pm 0.006$ & $5.29 \pm 1.90$ & $0.26 \pm 0.10$ \\
Day 5 (Oct 8) & $0.32^{+0.01}_{-0.01}$ & $1.83^{+0.01}_{-0.01}$ & $2.28^{+0.04}_{-0.03}$ & $0.56^{+0.02}_{-0.02}$ & $41.1 \pm 8.1$ & $108.0 \pm 21.2$ & $6.56^{+0.05}_{-0.05}$ & $0.28^{+0.06}_{-0.05}$ & $0.63^{+0.12}_{-0.11}$ & $1.14\ (294.4/257)$ & $13.0 \pm 0.3$ & $3.0 \pm 0.1$ & $17.91 \pm 0.02$ & $0.066 \pm 0.008$ & $5.38 \pm 1.94$ & $0.27 \pm 0.10$ \\
Day 6 (Oct 9) & $0.33^{+0.01}_{-0.01}$ & $1.88^{+0.01}_{-0.01}$ & $2.34^{+0.03}_{-0.05}$ & $0.55^{+0.01}_{-0.03}$ & $42.3 \pm 8.4$ & $110.4 \pm 21.9$ & $6.59^{+0.03}_{-0.03}$ & $0.2^{+0.04}_{-0.03}$ & $0.63^{+0.09}_{-0.09}$ & $0.98\ (253.6/256)$ & $12.4 \pm 0.3$ & $2.8 \pm 0.1$ & $17.16 \pm 0.02$ & $0.066 \pm 0.007$ & $5.15 \pm 1.86$ & $0.26 \pm 0.09$ \\
Day 7 (Oct 10) & $0.33^{+0.01}_{-0.0}$ & $1.85^{+0.01}_{-0.01}$ & $2.37^{+0.04}_{-0.04}$ & $0.56^{+0.02}_{-0.03}$ & $40.0 \pm 7.6$ & $104.5 \pm 19.9$ & $6.49^{+0.04}_{-0.04}$ & $0.19^{+0.04}_{-0.03}$ & $0.50^{+0.08}_{-0.08}$ & $1.16\ (299.1/256)$ & $12.6 \pm 0.3$ & $2.9 \pm 0.1$ & $17.35 \pm 0.02$ & $0.052 \pm 0.006$ & $5.21 \pm 1.87$ & $0.26 \pm 0.09$ \\ \hline
\multicolumn{17}{c}{4U $1608-52$}                                                                                                                                                                                                                                                                                                                                                                                                                                                                                                                                                                                                                                           \\ \hline
Pre-superburst  &  $1.01^{+0.02}_{-0.02}$ & $1.96^{+0.02}_{-0.02}$ & $2.59^{+0.09}_{-0.07}$ & $0.62^{+0.04}_{-0.04}$ & 
$20.9 \pm 3.6$ & $46.4 \pm 8.0$ & $6.74^{+0.05}_{-0.04}$ & $0.13^{+0.06}_{-0.04}$ & $0.39^{+0.09}_{-0.09}$ & $1.04\ (250.2/241)$ & $11.1 \pm 0.5$ & $2.5 \pm 0.2$ & $17.51 \pm 0.03$ & $0.042 \pm 0.008$ & $2.15 \pm 0.40$ & $0.11 \pm 0.02$\\
Post-superburst &  $1.02^{+0.02}_{-0.01}$ & $2.23^{+0.08}_{-0.12}$ & $2.80^{+0.12}_{-0.26}$ & $0.72^{+0.04}_{-0.06}$ & $16.0 \pm 2.4$ & $34.9 \pm 5.3$ & $6.71^{+0.04}_{-0.04}$ & $0.12^{+0.05}_{-0.04}$ & $0.29^{+0.08}_{-0.07}$ & $1.00\ (244.9/246)$ & $8.9 \pm 1.0$ & $1.7 \pm 0.4$ & $15.82 \pm 0.03$ & $0.031 \pm 0.006$ & $1.94 \pm 0.36$ & $0.10 \pm 0.02$\\
Day 1 (Mar 8) & $1.05^{+0.04}_{-0.01}$ & $1.93^{+0.02}_{-0.01}$ & $2.39^{+0.06}_{-0.05}$ & $0.59^{+0.03}_{-0.06}$ & $26.4 \pm 5.3$ & $60.9 \pm 12.2$ & $6.73^{+0.05}_{-0.03}$ & $0.22^{+0.07}_{-0.06}$ & $0.55^{+0.13}_{-0.12}$ & $1.01\ (251.1/247)$ & $11.9 \pm 0.4$ & $2.6 \pm 0.2$ & $21.32 \pm 0.03$ & $0.059 \pm 0.009$ & $2.62 \pm 0.49$ & $0.13 \pm 0.02$ \\
Day 2 (Mar 9) & $1.03^{+0.01}_{-0.03}$ & $1.87^{+0.05}_{-0.01}$ & $2.52^{+0.07}_{-0.05}$ & $0.59^{+0.07}_{-0.03}$ & $22.7 \pm 4.0$ & $53.1 \pm 9.4$ & $6.66^{+0.05}_{-0.05}$ & $0.15^{+0.08}_{-0.06}$ & $0.40^{+0.14}_{-0.10}$ & $1.08\ (271.4/249)$ & $12.0 \pm 0.6$ & $2.9 \pm 0.3$ & $22.92 \pm 0.04$ & $0.043 \pm 0.008$ & $2.81 \pm 0.53$ & $0.14 \pm 0.03$ \\
Day 3 (Mar 10) & $1.04^{+0.02}_{-0.02}$ & $1.84^{+0.01}_{-0.01}$ & $2.48^{+0.06}_{-0.03}$ & $0.59^{+0.03}_{-0.05}$ & $25.9 \pm 4.3$ & $60.5 \pm 9.9$ & $6.71^{+0.03}_{-0.03}$ & $0.11^{+0.06}_{-0.02}$ & $0.39^{+0.08}_{-0.09}$ & $1.17\ (297.2/251)$ & $12.4 \pm 0.3$ & $3.0 \pm 0.2$ & $22.57 \pm 0.03$ & $0.042 \pm 0.007$ & $2.77 \pm 0.52$ & $0.14 \pm 0.03$ \\
Day 4 (Mar 17) & $1.03^{+0.01}_{-0.04}$ & $2.03^{+0.05}_{-0.02}$ & $2.59^{+0.17}_{-0.09}$ & $0.64^{+0.08}_{-0.02}$ & $17.5 \pm 2.5$ & $38.7 \pm 5.5$ & $6.63^{+0.05}_{-0.05}$ & $0.21^{+0.05}_{-0.04}$ & $0.40^{+0.10}_{-0.09}$ & $1.16\ (289.0/248)$ & $10.5 \pm 0.7$ & $2.2 \pm 0.3$ & $16.74 \pm 0.03$ & $0.043 \pm 0.007$ & $2.06 \pm 0.39$ & $0.10 \pm 0.02$ \\
Day 5 (Mar 18) & $1.01^{+0.01}_{-0.01}$ & $1.87^{+0.01}_{-0.01}$ & $2.45^{+0.05}_{-0.05}$ & $0.59^{+0.03}_{-0.05}$ & $23.5 \pm 5.6$ & $52.5 \pm 12.5$ & $6.68^{+0.04}_{-0.05}$ & $0.21^{+0.06}_{-0.05}$ & $0.50^{+0.10}_{-0.10}$ & $1.13\ (283.3/249)$ & $12.2 \pm 0.4$ & $2.9 \pm 0.2$ & $18.10 \pm 0.03$ & $0.053 \pm 0.007$ & $2.22 \pm 0.42$ & $0.11 \pm 0.02$ \\
Day 6 (Mar 19) & $1.03^{+0.01}_{-0.02}$ & $2.0^{+0.04}_{-0.01}$ & $2.64^{+0.08}_{-0.08}$ & $0.63^{+0.05}_{-0.02}$ & $19.1 \pm 2.9$ & $42.1 \pm 6.5$ & $6.7^{+0.04}_{-0.05}$ & $0.21^{+0.05}_{-0.04}$ & $0.48^{+0.09}_{-0.10}$ & $1.14\ (286.9/250)$ & $10.6 \pm 0.5$ & $2.3 \pm 0.2$ & $16.95 \pm 0.03$ & $0.046 \pm 0.007$ & $2.08 \pm 0.39$ & $0.10 \pm 0.02$ \\
Day 7 (Mar 20) & $1.01^{+0.01}_{-0.02}$ & $2.2^{+0.07}_{-0.03}$ & $2.75^{+0.27}_{-0.11}$ & $0.69^{+0.05}_{-0.03}$ & $16.3 \pm 2.5$ & $35.4 \pm 5.4$ & $6.69^{+0.05}_{-0.05}$ & $0.15$\tnote{$\|$} & $0.29^{+0.06}_{-0.06}$ & $1.09\ (268.7/245)$ & $9.1 \pm 0.8$ & $1.8 \pm 0.3$ & $15.74 \pm 0.03$ & $0.033 \pm 0.007$ & $1.93 \pm 0.36$ & $0.10 \pm 0.02$ \\
Day 8 (Mar 21) & $1.01^{+0.02}_{-0.02}$ & $2.06^{+0.02}_{-0.02}$ & $2.69^{+0.08}_{-0.09}$ & $0.63^{+0.04}_{-0.02}$ & $19.5 \pm 2.7$ & $42.3 \pm 5.8$ & $6.69^{+0.03}_{-0.02}$ & $0.11^{+0.03}_{-0.03}$ & $0.30^{+0.07}_{-0.05}$ & $1.05\ (263.4/249)$ & $10.1 \pm 0.4$ & $2.1 \pm 0.2$ & $15.36 \pm 0.02$ & $0.033 \pm 0.005$ & $1.89 \pm 0.35$ & $0.09 \pm 0.02$ \\
Day 9 (Mar 22) & $1.03^{+0.08}_{-0.01}$ & $1.88^{+0.01}_{-0.02}$ & $2.63^{+0.06}_{-0.07}$ & $0.58^{+0.02}_{-0.08}$ & $28.0 \pm 4.6$ & $62.1 \pm 10.3$ & $6.72^{+0.05}_{-0.05}$ & $0.18^{+0.05}_{-0.05}$ & $0.36^{+0.09}_{-0.04}$ & $1.19\ (300.2/250)$ & $11.6 \pm 0.4$ & $2.8 \pm 0.2$ & $17.46 \pm 0.03$ & $0.039 \pm 0.007$ & $2.15 \pm 0.40$ & $0.11 \pm 0.02$ \\
Day 10 (Mar 23) & $1.02^{+0.03}_{-0.01}$ & $1.9^{+0.02}_{-0.01}$ & $2.75^{+0.09}_{-0.08}$ & $0.56^{+0.03}_{-0.03}$ & $24.9 \pm 4.3$ & $54.8 \pm 9.5$ & $6.66^{+0.05}_{-0.05}$ & $0.16^{+0.05}_{-0.02}$ & $0.33^{+0.06}_{-0.08}$ & $0.95\ (241.5/253)$ & $11.2 \pm 0.4$ & $2.7 \pm 0.2$ & $16.61 \pm 0.02$ & $0.035 \pm 0.006$ & $2.04 \pm 0.38$ & $0.10 \pm 0.02$ \\
Day 11 (Mar 24) & $1.05^{+0.04}_{-0.01}$ & $2.2^{+0.03}_{-0.02}$ & $3.05^{+0.18}_{-0.13}$ & $0.61^{+0.01}_{-0.05}$ & $22.1 \pm 3.1$ & $46.6 \pm 6.6$ & $6.66^{+0.04}_{-0.04}$ & $0.16^{+0.05}_{-0.04}$ & $0.32^{+0.06}_{-0.06}$ & $0.88\ (218.6/247)$ & $8.6 \pm 0.5$ & $1.8 \pm 0.2$ & $13.52 \pm 0.02$ & $0.034 \pm 0.005$ & $1.66 \pm 0.31$ & $0.08 \pm 0.02$ \\
Day 12 (Mar 25) & $1.0^{+0.0}_{-0.03}$ & $2.21^{+0.05}_{-0.03}$ & $3.06^{+0.24}_{-0.2}$ & $0.7^{+0.04}_{-0.04}$ & $15.4 \pm 2.2$ & $32.7 \pm 4.7$ & $6.72^{+0.06}_{-0.03}$ & $0.26^{+0.07}_{-0.06}$ & $0.44^{+0.13}_{-0.13}$ & $1.14\ (283.3/247)$ & $8.5 \pm 0.8$ & $1.7 \pm 0.3$ & $13.65 \pm 0.02$ & $0.047 \pm 0.006$ & $1.68 \pm 0.31$ & $0.08 \pm 0.02$ \\
Day 13 (Mar 26) & $1.05^{+0.02}_{-0.01}$ & $1.9^{+0.02}_{-0.01}$ & $2.61^{+0.08}_{-0.03}$ & $0.58^{+0.03}_{-0.04}$ & $22.9 \pm 3.7$ & $50.0 \pm 8.1$ & $6.64^{+0.05}_{-0.05}$ & $0.16^{+0.06}_{-0.05}$ & $0.32^{+0.08}_{-0.08}$ & $1.01\ (254.3/250)$ & $11.5 \pm 0.3$ & $2.7 \pm 0.2$ & $16.08 \pm 0.02$ & $0.034 \pm 0.006$ & $1.98 \pm 0.37$ & $0.10 \pm 0.02$ \\ \hline
\end{tabular}%
\egroup
\begin{tablenotes} 
 			\item[*] $\times 10^{22}$ cm$^{-2}$
            \item[\textdagger] $\times 10^{-2}$ ph~cm$^{-2}$~s$^{-1}$
            \item[\textdaggerdbl] $\times 10^{-9}$~erg~s$^{-1}$~cm$^{-2}$, unabsorbed, $0.8 - 15$ keV
            \item[\S]  $\times 10^{37}$~erg~s$^{-1}$  
            \item[$\|$] Could not be constrained, so frozen to average of other observations   
\end{tablenotes}
\end{threeparttable}
\end{table}

\end{landscape}


\begin{figure*}
\centering
\begin{subfigure}{0.49\textwidth}
\includegraphics[scale=0.57, trim=0cm 0cm 0cm 0cm, clip=true]{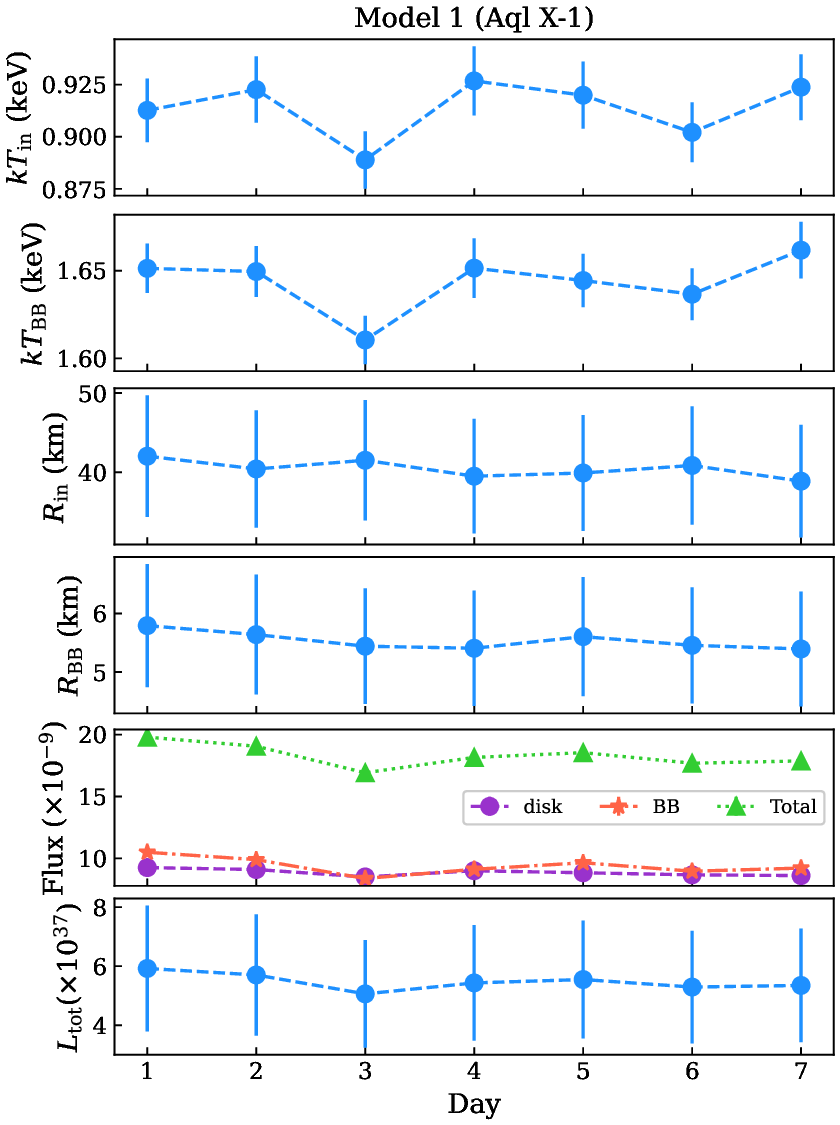}
\caption{}
\end{subfigure}\hfill
\begin{subfigure}{0.49\textwidth}
\includegraphics[scale=0.57, trim=0cm 0cm 0cm 0cm, clip=true]{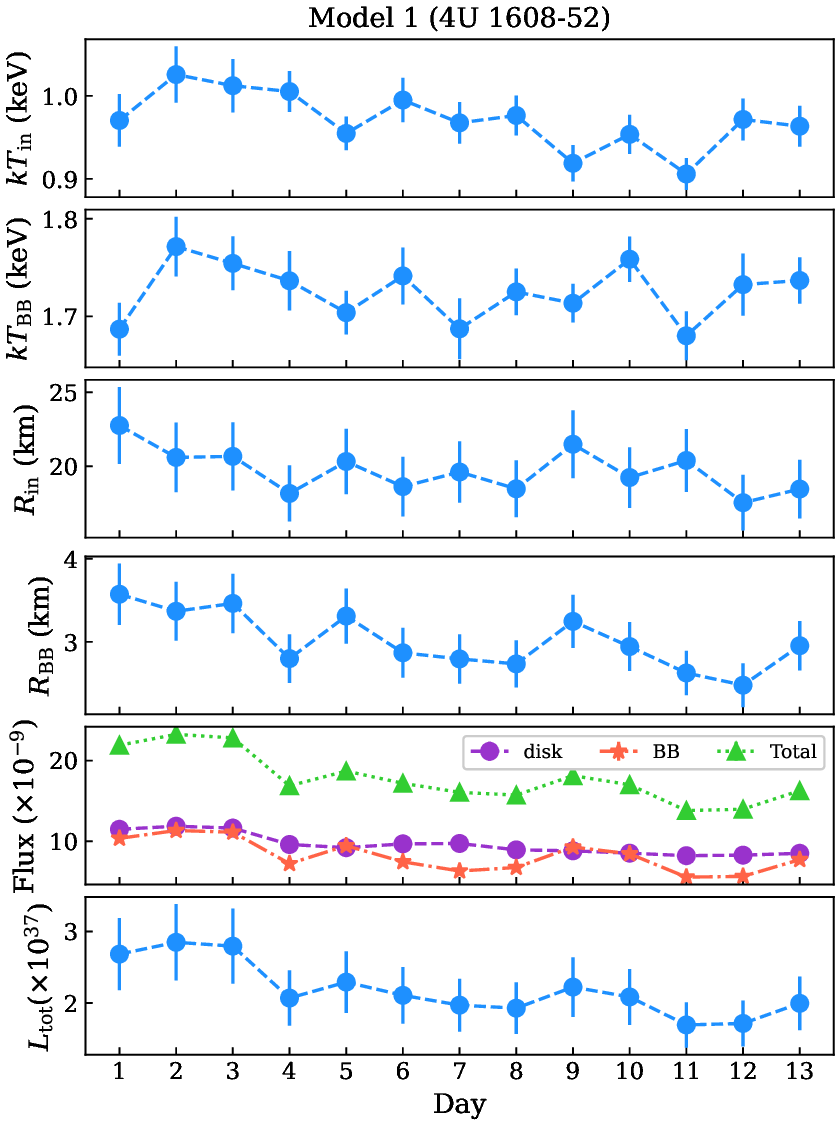}
\caption{}
\end{subfigure}  
\caption{Best fit parameters using model~1 to day-wise spectra of (a) \aql, and (b) \fouru. Fluxes and luminosities are unabsorbed values in $0.8-15$~keV, and are plotted in units of $10^{-9}$~erg~s$^{-1}$~cm$^{-2}$ and erg~s$^{-1}$ respectively}
\label{fig:daywiseparams_model1}
\end{figure*} 

\begin{figure*}
\centering
\begin{subfigure}{0.49\textwidth}
\includegraphics[scale=0.57, trim=0cm 0cm 0cm 0cm, clip=true]{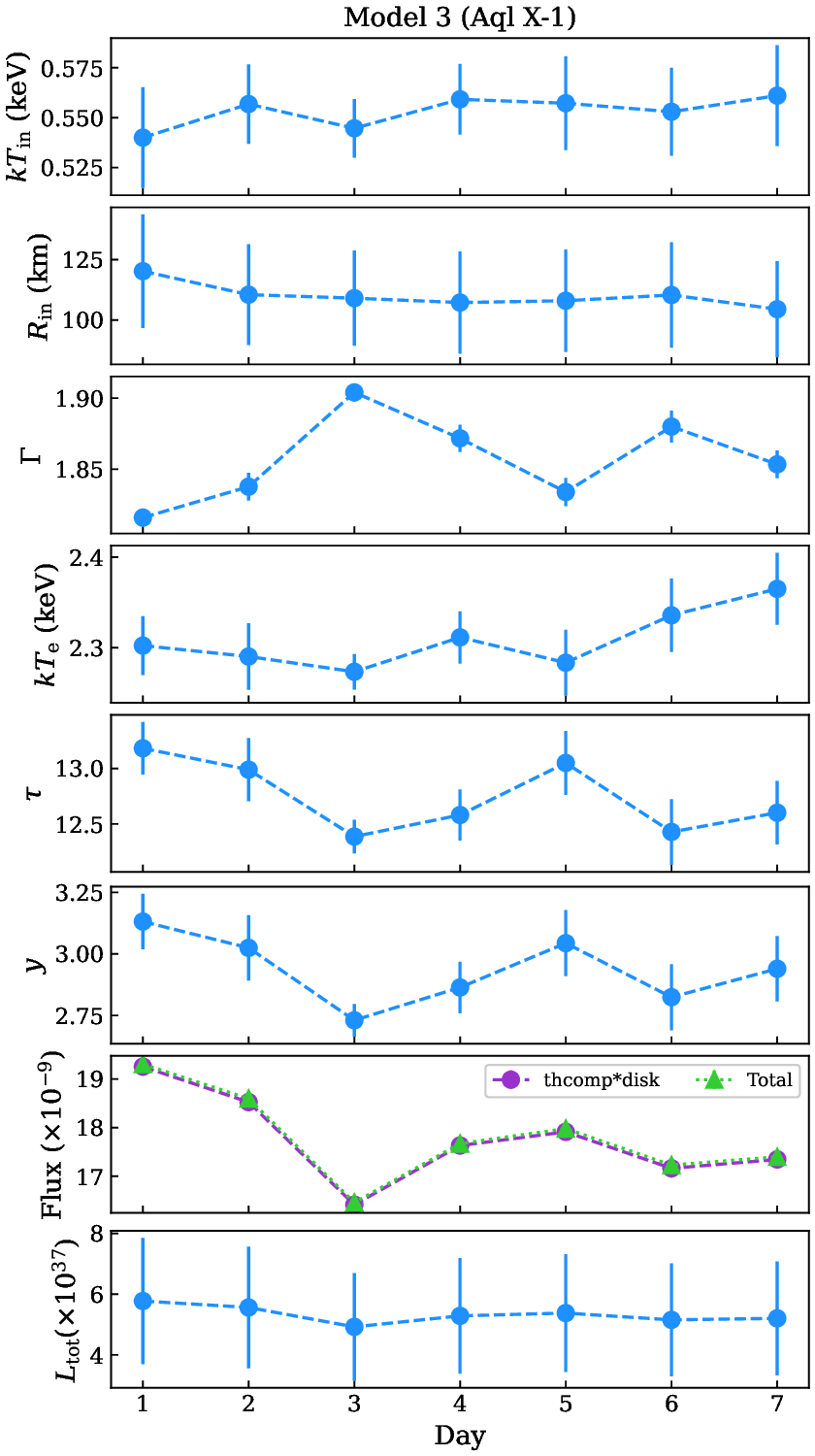}
\caption{}
\end{subfigure}\hfill
\begin{subfigure}{0.49\textwidth}
\includegraphics[scale=0.57, trim=0cm 0cm 0cm 0cm, clip=true]{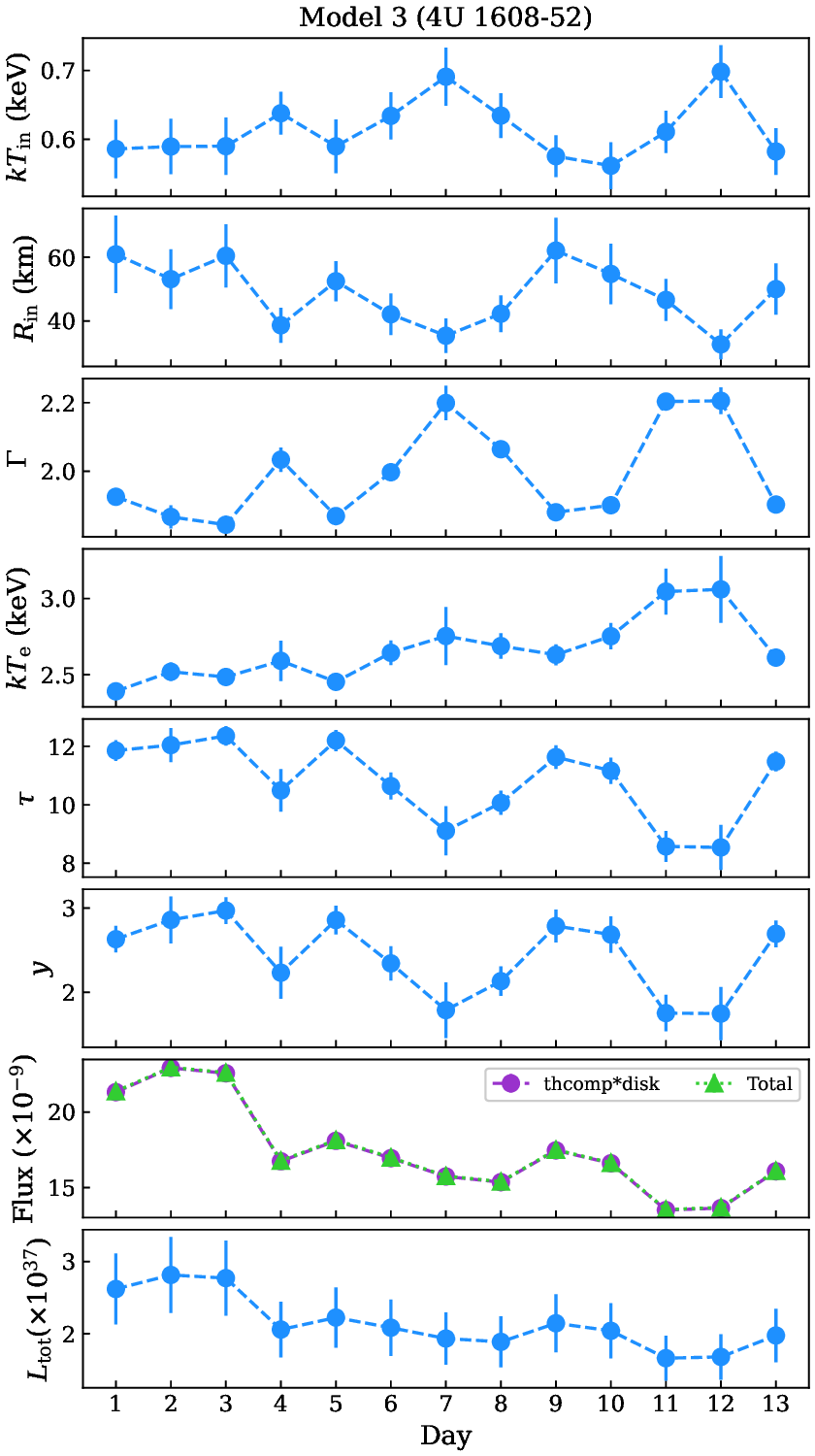}
\caption{}
\end{subfigure}  
\caption{Same as Figure~\ref{fig:daywiseparams_model1} but using model~3.}
\label{fig:daywiseparams_model3}
\end{figure*} 

\subsubsection*{\aql}
The best-fit hydrogen column density ($N_H$) using both the models is found to range from $(0.32-0.39)\times 10^{22}$~cm$^2$. This value is consistent with the value of Galactic $N_H$ \citep{hi4pi2016} along the line of sight to \aql, as well as those obtained by some authors (\citealt{raichur2011, abdelfatah2021}). However, slightly higher values of $N_H$ are also found in literature (\citealt{campana2014, keek2018, bult2018, putha2024}).

The variation of spectral parameters using model~1 are shown in Figure~\ref{fig:daywiseparams_model1}a. Using this model, the disk component does not vary much with time, with inner disk temperature $\sim 0.9$~keV. Assuming a distance of $5\pm 0.9$~kpc and orbital inclination of 36$^{\circ}$, the color-corrected inner radius of the accretion disk is found to be $\sim 40$~km. The blackbody temperature remains nearly constant at $\sim 1.65$~keV, with blackbody radius $\sim 5.5$~km, implying that the effective emission area of the blackbody is smaller than the entire NS surface. The unabsorbed $0.8-15$~keV flux is between $(1.7-2.0)\times 10^{-8}$~ergs~s$^{-1}$~cm$^{-2}$, mimicking the slight variations in the light curve (bottom panel of Figure~\ref{fig:fullaql}). The contribution of the disk and blackbody components to the total flux are comparable throughout the observations. 

Using model~3 (Figure~\ref{fig:daywiseparams_model3}a), however, the inner disk temperature is estimated to be slightly lower ($\sim 0.55$~keV), whereas the inner radius of the accretion disk is $\sim 3$~times higher ($\sim 100-120$~km). The photon index does not show any systematic trends, with an average value of $1.86$, whereas the electron temperature shows a weak increasing trend. The average optical depth ($\tau$) and Compton y-parameter are computed to be $\sim 12.7$ and $2.9$ respectively during the observation.

The energy of the fitted Gaussian line lies between $6.5-6.7$~keV, with line width $\sigma\sim 0.1-0.2$~keV. Using both the models, the Gaussian line flux is more than two orders of magnitude lower ($\lesssim 0.5\%$) than the total flux. 

\subsubsection*{\fouru}
Using both the models, $N_H$ is well-constrained in the range $(1.0-1.1)\times 10^{22}$~cm$^2$, which is consistent with the values found in literature (\citealt{guver2010, armaspadilla2017, jaisawal2019, chen2022}). 

For model~1, the inner disk temperature remains within $\sim 0.9-1.0$~keV. The inner disk radius (calculated assuming a distance of $3.2\pm 0.3$~kpc and inclination of $35^{\circ}$) does not show much variation, and is found to be $\sim 20$~km. Similar to \aql, the blackbody temperature remains $\sim 1.7$~keV, whereas the radius decreases from $\sim 3.6$~km to $\sim 2.5$~km. This is much smaller than the typical NS radius, indicating that the blackbody emission may come from parts of the NS surface. The flux of both the components are comparable on most days, with the disk component being upto a factor of 1.5 stronger than the blackbody component on others.

Fitting the spectra with model~3 result in a lower estimate of the inner disk temperature ($\sim 0.6-0.7$~keV), and higher estimate of the inner disk radius ($\sim 30-60$~km). The photon index of the Comptonization component fluctuates between $1.8-2.2$, without any clear day-to-day trend, whereas the electron temperature steadily increases from $2.4-3.1$~keV (barring day~13, when it drops to 2.6~keV). The $\tau$ and y-parameter both follow a similar trend, exhibiting a weak decreasing trend overall, but with significant day-to-day fluctuations. The high optical depth ($8.5-12.4$) and Compton y-parameter ($1.7-3.0$) indicates that the Corona is optically thick with multiple scattering of the photons.

The evolution of various parameters using these two models for this source are plotted in Figures~\ref{fig:daywiseparams_model1}b and \ref{fig:daywiseparams_model3}b respectively.

\subsubsection{Superburst of \fouru}
\label{subsec:spec_sb}

\begin{table*}
\centering
\begin{threeparttable}
\caption{Best fit spectral parameters for the six coarse superburst segments, with an $f_a$ scaling to model~3, and an additional blackbody component. Pre-burst parameters are frozen. Average count rate quoted is for 11 SCDs.}

\label{tab:superburst_params}
\bgroup
\def\arraystretch{1.5}
\begin{tabular}{ccccccccccc}
\hline
Interval & T$_{\mathrm{exp}}$ (s) & Avg rate (cps) & $f_a$                  & $kT_{BB}$              & $R_{BB}$               & Total flux\tnote{*}          & Blackbody flux\tnote{*} & Luminosity \tnote{\textdagger}     & L/L$_{\mathrm{Edd}}$ & $\chi^2_{\mathrm{red}}\,(\chi^2/dof)$ \\ \hline
t1       & 600                    & 78.3           & $0.96^{+0.02}_{-0.02}$ & $1.91^{+0.02}_{-0.02}$ & $4.83\pm 0.49$ &  $48.28\pm 0.51$ & $31.29 \pm 0.41$ &   $5.92\pm 1.11$ & $0.3\pm 0.06$ & 1.19 (242.4/204)                      \\
t2       & 992                    & 61.7           & $0.91^{+0.02}_{-0.02}$   & $1.48^{+0.01}_{-0.01}$ & $5.88\pm 0.59$ &  $33.21\pm 0.4$ & $17.13 \pm 0.29$ &   $4.07\pm 0.76$ & $0.2\pm 0.04$ &  1.48 (308.1/208)                      \\
t3       & 592                    & 46.2           & $0.93^{+0.03}_{-0.03}$ & $1.27^{+0.02}_{-0.02}$ & $5.27\pm 0.67$ &   $23.85\pm 0.55$ & $7.43 \pm 0.37$ &   $2.92\pm 0.55$ & $0.15\pm 0.03$ &  1.36 (261.8/193)                      \\
t4       & 174                    & 38.5           &  $0.88^{+0.06}_{-0.06}$  & $0.94^{+0.04}_{-0.04}$ & $6.95\pm 1.46$ &  $19.55\pm 1.25$ & $3.95 \pm 0.78$ &   $2.39\pm 0.47$ & $0.12\pm 0.02$ &  0.88 (138.1/156)                      \\
t5       & 598                    & 36.6           & $0.88^{+0.03}_{-0.03}$ & $1.04^{+0.03}_{-0.03}$ & $5.21\pm 0.89$ &  $18.74\pm 0.64$ & $3.23 \pm 0.41$ &   $2.3\pm 0.44$ & $0.11\pm 0.02$ &  1.06 (200.5/190)                      \\
t6       & 1541                   & 33.0           & $0.88^{+0.02}_{-0.02}$ & $0.80^{+0.03}_{-0.03}$ & $6.08\pm 0.92$ &   $17.05\pm 0.33$ & $1.56 \pm 0.2$ &   $2.09\pm 0.39$ & $0.1\pm 0.02$ &  1.24 (250.4/202)                      \\ \hline
\end{tabular}
\egroup
\begin{tablenotes} 
            \item[*] Unabsorbed flux in $0.8-15$~keV, in units of $\times 10^{-9}$~erg~s$^{-1}$~cm$^{-2}$
            \item[\textdagger] Total luminiosity in units of $\times 10^{37}$~erg~s$^{-1}$
\end{tablenotes}
\end{threeparttable}
\end{table*}

To understand the spectral evolution of the source during the superburst, we carried out time-resolved spectroscopy covering the event. We extracted six spectra corresponding to the six intervals marked 1 to 6 in Figure~\ref{fig:superburst}, along with the pre-burst and post-burst spectra. In addition, we also carried out time-resolved analysis in finer intervals, to track the complete evolution of the source from the precursor burst to the superburst decay. For this analysis, we extracted spectra with durations ranging from 30~seconds to 240~seconds\footnote{Durations were gradually increased as count rate decayed.}, and included the duration of the precursor burst decay\footnote{The three spectra covering the precursor burst have exposures 0.7~s, 2~s and 6~s respectively.}, and the time preceding the superburst rise as well. This corresponds to the intervals 1 to 6, as well as the duration following `pre-burst' and preceding interval~1. As earlier, we applied optimal binning to the spectra, and restricted the fitting range from $0.8-10$~keV.

First we fitted the pre-burst spectrum with models~1 and 3, and the best fit parameters are summarized in Tables~\ref{tab:model1} and Table~\ref{tab:model3}. The parameters are broadly consistent with the day-wise fit parameters within statistical uncertainties, implying no detectable spectral change preceding the superburst. In the following, we adopt model~3 as our best fit model of persistent emission. 

\begin{figure*}
\centering 
\begin{subfigure}{0.24\textwidth}
\centering
\includegraphics[scale=0.31]{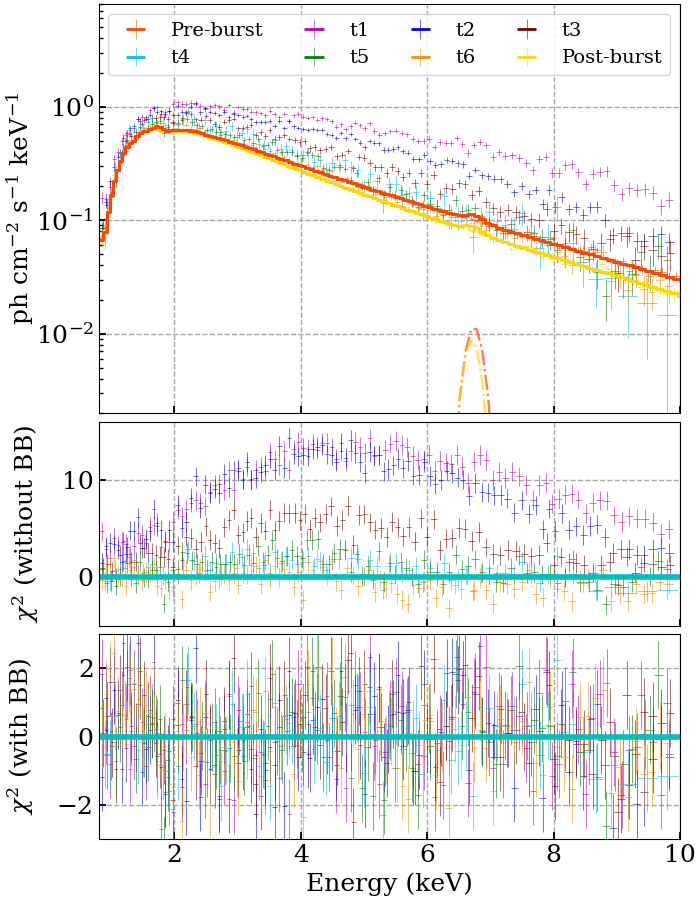}
\caption{}
\end{subfigure}\hfill
\begin{subfigure}{0.74\textwidth}
\centering
\includegraphics[scale=0.29, trim = 0cm 0.0cm 0.5cm 0cm, clip=true]{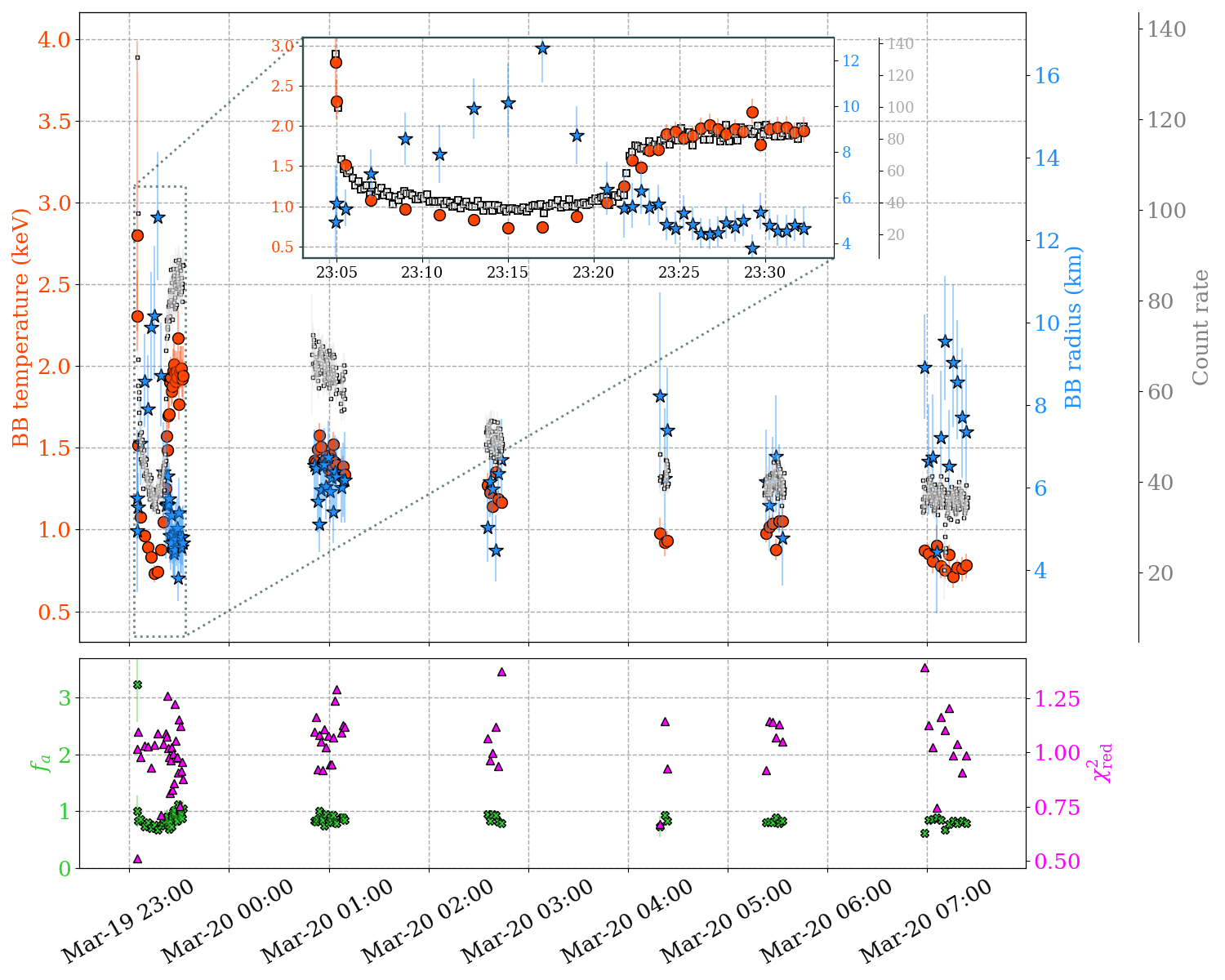}
\caption{}
\end{subfigure} 
\caption{(a) (Top) Unfolded spectra of the six superburst segments (t1 to t6), fitted with absorbed \texttt{fa*thcomp*diskbb + bbodyrad}. The pre-burst and post-burst spectra are also shown, along with the best fit, using model~3. (Middle) Residuals of spectral fits to t1 to t6 without the blackbody component, and (bottom) after including the blackbody component. For clarity, only $2^{\circ}\times 2^{\circ}$ spectra are shown. (b) Best fit parameters of time-resolved spectroscopy of the superburst in fine bins: (Top) Blackbody temperature and radius are plotted on the left (red circles) and right (blue stars) axes respectively. Count rate is overplotted on the far right axis (grey squares). The inset shows a zoomed in portion of the precursor and beginning of superburst. (Bottom) $f_a$ and $\chi^2_{\mathrm{red}}$ are plotted on the left (green cross) and right (magenta triangle) axes respectively.}
\label{fig:superburst_spec}
\end{figure*}

For a broad look at the spectral evolution, we first tried to fit the coarsely binned spectra (six intervals) with model~3 . There are large residuals, implying the spectra cannot be described with this model alone (Figre~\ref{fig:superburst_spec}). Next, we added a blackbody component to the model, and employed two approaches while fitting. In the first, we froze the pre-burst fit parameters, leaving the temperature and normalization of the blackbody component to vary freely. In the second, we allowed the normalization of the pre-burst emission to vary during the burst, This is the so-called `$f_a$' method, where $f_a$ is a free scaling factor that can be used to account for the variation of persistent emission.

We find that the $f_a$ method results in improved fits as compared to the fits without $f_a$. The best fit spectral parameters ($f_a$, blackbody temperature, and radius) are summarized in Table~\ref{tab:superburst_params}. Figure~\ref{fig:superburst_spec}a (top panel) shows these six spectra, along with the pre-burst and post-burst best fits. In the middle panel, the residuals are shown when fitting the superburst spectra with the pre-burst model, without adding a blackbody component. In the bottom panel, the same is shown after including a blackbody component, along with a scaling of the persistent emission. Thus, adding a blackbody component results in improved fits for all the intervals. 

The blackbody temperature steadily declines from 1.9~keV to 0.8~keV over the six time intervals\footnote{With the exception of t4. The temperatures in t4 and t5 are consistent within statistical uncertainties.}. The blackbody flux also decreases from $3.1\times 10^{-8}$~erg~s$^{-1}$~cm$^{-2}$ to $1.6\times 10^{-9}$~erg~s$^{-1}$~cm$^{-2}$. We note that during type-I bursts, the persistent emission is typically found to increase, giving $f_a > 1$ (e.g. \citealt{worpel2013, jaisawal2019, guver2022}). This effect is usually attributed to the Poynting-Robertson drag \citep{walker1992}. However, we find that during the superburst, the $f_a$ is always less than 1, indicating a suppression of the persistent emission. 

Since the $f_a$ is just a scaling factor, it cannot account for more complex phenomena such as change in the spectral shape of the emission. To test if the accretion disk itself is affected by the superburst, we removed the $f_a$ and freed the disk parameters (temperature and normalization) while fitting the superburst spectra. A blackbody component, as earlier, is required for good fits. We found that $kT_{in}$ is systematically lower than the pre-burst value of $0.62$~keV, accompanied by a higher inner disk radius. However, due to insufficient statistics, the disk parameters are not well-constrained, hence we do not explore this method further.

Next, to resolve finer variations in the spectral properties during the superburst, we proceed to fit the finely spaced spectra, with the same model i.e. $f_a\times$~pre-burst emission + blackbody. This model produces good fits to all the spectra. The evolution of the blackbody temperature and corresponding radius during the superburst is shown in the top panel of Figure~\ref{fig:superburst_spec}(b). For comparison, the light curve of the same duration is also overplotted. The bottom panel shows the obtained $f_a$ and goodness-of-fit ($\chi^2_{\mathrm{red}}$).

At the peak of the precursor burst, the temperature rises to $\sim 2.8$~keV, accompanied by an $f_a$ value of 3.2 and blackbody radius of 5~km. In the interval between the precursor decay and superburst, the blackbody temperature slowly decreases, $f_a$ decreases to $\sim 0.7$, and the radius increases upto 12.6~km. Subsequently, as the source brightens again to the superburst, the temperature slowly climbs again and flattens out at $\sim 2$~keV (see inset in the top panel of Figure~\ref{fig:superburst_spec}b). The temperature rise is exactly followed by a radius decay to $\sim 4.7$~km. During this period, $f_a$ remains $\lesssim 1$. Then in the five subsequent orbits, the blackbody temperature declines as the superburst decays, accompanied by a weak increase in radius. Note that the parameter values quoted in  Table~\ref{tab:superburst_params} are for the coarsely binned spectra, hence they represent an average of the fine spectra in the respective segments. The evolution of the blackbody flux during the superburst is further discussed in Section~\ref{sec:disc}.

Finally, we fit the post-burst spectra (duration marked in Figure~\ref{fig:superburst}) to compare with the pre-burst and burst durations. The best fit parameters are listed in Tables~\ref{tab:model1} and \ref{tab:model3}. Compared to the pre-burst spectral fit using model~3, the post-burst spectra is slightly softer ($\Gamma=2.23$ vs 1.96), as also evident from Figure~\ref{fig:superburst_spec}a. This can be attributed to the influx of soft photons during the superburst, before relaxing to the pre-burst value over the next two to three days (day~$8-10$). 

\subsubsection{Type-I X-ray burst}
\label{subsec:spec_bur}
Time-resolved spectroscopy can help understand how the spectral properties of the source change during a Type-I burst. Unfortunately, due to the small effective area of the instrument, we do not have enough statistics to carry out spectroscopy in finely resolved time bins. Especially, signatures during the initial phase of the burst can reveal whether it exhibits Photospheric Radius Expansion (PRE), and these studies could not be carried out for the three bursts observed (AqB1, 4UB1, and 4UB2).

Nevertheless, since bursts are bright phenomena, it is possible to carry out time resolved spectroscopy in coarse bins. We divided each burst duration into three segments, roughly covering the peak, the intermediate stage and the low phases ($t \gtrsim \tau$) of the burst, with approximate durations of $2-3$~s, $5-6$~s and $40-50$~s respectively. For each burst, we also extracted a pre-burst spectrum, and fitted it with model~3. For fitting the bursts spectra, we again use the $f_a$ method for scaling the persistent emission, and include an additional blackbody component. 

The best-fit spectral parameters ($f_a$, blackbody temperature, and blackbody radius) for each segment, designated `high', `mid' and `low', are summarized in Table~\ref{tab:burst_params}. For all three bursts, $f_a > 1$ at least near the burst peak, with $f_a$ reaching 3.2 for the precursor burst. All the bursts show a cooling trend over the three bins, as expected during type-I bursts. The peak flux ($F_{peak}$) of AqB1 is found to be $5.9\times 10^{-8}$~erg~s$^{-1}$~cm$^{-2}$, which corresponds to $\sim 0.9L_{Edd}$. 4UB1 and 4UB2 have peak fluxes of $1.8\times 10^{-7}$~erg~s$^{-1}$~cm$^{-2}$ and $1.2\times 10^{-7}$~erg~s$^{-1}$~cm$^{-2}$, with the former reaching the Eddington limit, and the latter $\sim 0.7L_{Edd}$. We note here that the touchdown flux of \fouru, as estimated from PRE bursts, is reported as $18.5\times 10^{-8}$~erg~s$^{-1}$~cm$^{-2}$ by \cite{ozel2016}. This provides further evidence that 4UB1 was indeed Eddington limited. Moreover, this also indicates that the distance to \fouru, taken as $3.2\pm 0.3$~kpc in this work, is a conservative upper limit and higher distances to the source are unlikely as it will yield super-Eddington luminosities for this burst.

The fluence gives a measure of total energy emitted per unit area, integrated over the burst duration. Each burst was fitted with a FRED profile (Section~\ref{subsec:tim_bur}). To compute the burst duration, the start time was chosen as the time when the flux exceeds 10\% of the peak flux. The $T_{decay}$, defined as the time at which the flux declines to $<10\%$ of the peak value, was chosen as the end time. The spectra in this duration was fitted and the flux was calculated. After subtracting the pre-burst flux, the total burst fluence $E_b$ was calculated, which is given in Table~\ref{tab:burst_params}. The total burst energy release $E_{rad}$, estimated from the fluence is also listed in Table~\ref{tab:burst_params}.

We estimated the column depth $y_{ign}$ at which the burst is ignited using the relation (see e.g. \citealt{galloway2008})
\begin{equation}
y_{ign}=\frac{4\pi d^2 E_b(1+z)}{4\pi R_{\mathrm{NS}}^2 Q_{\mathrm{nuc}}}
\end{equation}
where $(1+z)$ represents the gravitational redshift correction, $d$ is the distance to the source, $R_{\mathrm{NS}}$ is the neutron star radius, and $Q_{\mathrm{nuc}}$ is the energy released per nucleon. For a canonical neutron star with $M_{\mathrm{NS}}=1.4 M_{\odot}$ and $R_{\mathrm{NS}}=10$~km, $1+z=(1-2GM_{\mathrm{NS}}/R_{\mathrm{NS}}c^2)^{-1/2}\sim 1.31$, and taking $Q_{\mathrm{nuc}}=4.4$~MeV~nucleon$^{-1}$ for material with solar abundance, we find $y_{ign}$ to lie in the range $(2-3)\times 10^7$~g~cm$^{-2}$.

\begin{table}
\centering
\begin{threeparttable}
\caption{Timing and spectral parameters of the three type-I bursts observed from \aql and \fouru, and of the superburst}
\label{tab:burst_params}
\bgroup
\def\arraystretch{1.4}
\setlength{\tabcolsep}{1.8pt}
\begin{tabular}{lcccc}
\hline
Parameter  & \textbf{Burst 1} & \textbf{Burst2} & \textbf{Burst 3} & \textbf{Superburst} \\ \hline
ID                       & AqB1          & 4UB1          & 4UB2    &  -         \\
t$_{\mathrm{start}}$ (MJD)     &  60590.44794  & 60753.96173 &            60759.38362  &  60753.96875 \\
T$_{\mathrm{rise}}$ (s)     &  0.7\tnote{*}     & $1.4\pm 0.2$   & $1.7\pm 0.1$     &   $\sim 820$       \\
$\tau$ (s)  & $9.4\pm 0.7$  & $6.0\pm 0.4$  & $5.7\pm 0.2$                   & $10986\pm 259$ \\
T$_{\mathrm{decay}}$  (s)  &  $21.7\pm 1.6$   &   $13.8\pm 1.0$  &  $13.1\pm 0.5$         &   $25297\pm 597$   \\
Preburst rate\tnote{\textdagger} & $70.0\pm 1.6$ & $25.4\pm 3.2$   & $32.0\pm 1.3$  & $35.0\pm 0.7$  \\
Postburst rate\tnote{\textdagger} & \tnote{\textdaggerdbl}   &   $56.5\pm 1.3$    &   \tnote{\textdaggerdbl}    &  $31.2\pm 0.2$  \\ 
$F_{peak}$ ($\times 10^{-8}$)\tnote{\S} & $5.9\pm 0.2$ & $18.3\pm 1.7$  & $11.5\pm 0.5$   & $5.06\pm 0.03$  \\
$L_{peak}/L_{Edd}$ & $0.9\pm 0.3$ & $1.1\pm 0.2$  &  $0.7\pm 0.1$ &  $0.3\pm 0.1$ \\
$f_{a,high}$     &  $1.25\pm 0.15$  &  $3.23\pm 0.64$  &        $1.70\pm 0.46$   & \tnote{\#} \\
$f_{a,mid}$     &  $1.12\pm 0.11$  &  $1.00\pm 0.28$  & $0.66\pm 0.32$          &  \tnote{\#} \\
$f_{a,low}$     &  $0.97\pm 0.05$  &  $0.82\pm 0.08$   &         $0.86\pm 0.17$   &  \tnote{\#}\\
$kT_{BB,high}$ (keV)    &  $2.45\pm 0.39$  &  $2.80\pm 0.58$   &  $2.04\pm 0.20$        &   \tnote{\#}\\
$kT_{BB,mid}$ (keV)      &  $2.57\pm 0.63$  &  $2.30\pm 0.25$  &   $1.62\pm 0.13$       &   \tnote{\#}\\
$kT_{BB,low}$ (keV)      &  $1.34\pm 0.23$    &  $1.52\pm 0.07$  & $1.14\pm 0.11$   &  \tnote{\#}\\
$R_{BB,high}$ (km)      &    $5.3\pm 1.6$   & $5.0\pm 2.0$   &  $7.3\pm 1.4$         &  \tnote{\#}\\
$R_{BB,mid}$ (km)      &     $3.7\pm 1.4$    &  $5.7\pm 1.1$   &  $8.0\pm 1.5$        &   \tnote{\#}\\
$R_{BB,low}$ (km)      &  $4.2\pm 1.9$     &  $5.5\pm 0.8$   &  $6.4\pm 1.8$   &  \tnote{\#}\\
\multirow{2}{*}{$E_b$ ($\times 10^{-7}$)\tnote{\S}}  
& \multirow{2}{*}{$4.3\pm 0.2$}
& \multirow{2}{*}{$9.8\pm 0.3$}
& \multirow{2}{*}{$7.0\pm 0.2$}
& $3016\pm 113$\tnote{$\alpha$} \\
&  &  &  & $18247\pm 648$\tnote{$\alpha$} \\
\multirow{2}{*}{$E_{rad}$ ($\times 10^{39}$)\tnote{$\|$}} 
& \multirow{2}{*}{$2.2\pm 0.8$}
& \multirow{2}{*}{$2.0\pm 0.4$}
& \multirow{2}{*}{$1.5\pm 0.3$}
& $630\pm 120$\tnote{$\alpha$} \\
& &  &  & $3812\pm 135$\tnote{$\alpha$} \\
\multirow{2}{*}{$y_{ign}$ ($\times 10^8$~g~cm$^{-2}$)} 
& \multirow{2}{*}{$0.31\pm 0.11$}
& \multirow{2}{*}{$0.30\pm 0.06$}
& \multirow{2}{*}{$0.21\pm 0.04$}
& $4009\pm 766$\tnote{$\alpha$} \\
&  &  &  & $14844\pm 493$\tnote{$\alpha$} \\

\hline
\end{tabular}
\egroup
\begin{tablenotes} 
            \item[*] Error could not be constrained
            \item[\textdagger] In cts~s$^{-1}$, using fifteen detectors for AqB1, and eleven for 4UB1, 4UB2, and superburst
            \item[\textdaggerdbl] Constrained fit, same as pre-burst rate
            \item[\S] Unabsorbed values in $0.8-15$~keV, in units of erg~s$^{-1}$~cm$^{-2}$ (flux) and erg~cm$^{-2}$ (fluence)
            \item[$\|$] Total X-ray energy released (in erg), computed at the NS surface, using the respective fluence and distance, taking $1+z=1.31$
            \item[$\!$\#] Spectral fit parameters $f_a$, $kT_{BB}$, and $R_{BB}$ of the superburst are given in Table~\ref{tab:superburst_params}.
            \item[$\alpha$] The first value is obtained by fitting the X-ray flux, the second is from fitting the cooling models by \cite{cumming2004} (see Section~\ref{subsec:disc_sb}).
        \end{tablenotes}
    \end{threeparttable}
\end{table}

\section{Discussion}
\label{sec:disc}

\subsection{Persistent spectral parameters}
We have carried out a detailed spectral analysis of \aql and \fouru using two different model prescriptions for their emission (model~1: disk plus blackbody emission, and model~3: Comptonized disk emission). We find that both these models give statistically acceptable and equivalent fits to the persistent duration spectra. An important finding is that when using a Comptonized disk prescription, there is no evidence for an additional blackbody component. Including of this component either worsens the fit slightly, or the improvement is not statistically significant, with high p-value ($\sim 15-20\%$) of improvement by chance (low F-statistic value). This indicates that the Comptonizing region intercepts most of the soft photons, so the thermal component is not visible directly or is weaker compared to the Comptonized component. 

We also tried to fit the continuum emission with alternate models such as \texttt{thcomp*bbodyrad} and \texttt{comptb}. In the former, although the fits are of good statistical quality, we obtain unphysically low values of blackbody temperature ($\sim 0.4-0.5$~keV). For the latter, the fits are equivalent to model~2 (\texttt{nthComp}) and the Comptonization model adopted by us (model~3 \texttt{thcomp}). 

\subsubsection*{\aql}
Using model~1, the spectra can be described with a multicolor disk of inner temperature $\sim 0.9$~keV, along with a $\sim 1.6$~keV blackbody component for emission from the neutron star surface/boundary layer. Using model~3, the disk photons are treated as seed photons, which are Comptonized by the coronal electrons. In this prescription, soft disk photons of inner temperature $\sim 0.5$~keV undergo inverse Compton scattering by an optically thick corona. 

\cite{maitra2004} fitted the RXTE/PCA spectra of \aql acquired during the 2000 outburst using the model combination of disk blackbody and a power-law, with the power law component interpreted as emission from the corona. As the source transitioned from the hard to soft state, the disk temperature increased from $\sim 0.5$~keV to $\sim 2$~keV accompanied by an increase in the disk radius, and the photon index softened. \cite{putha2024} carried out a detailed spectroscopy of the 2019 and 2020 outbursts of \aql using NICER observations, covering both the hard and soft states. In particular, they found that the soft state spectrum can be well described by a partially Comptonized accretion disk with an inner temperature of $\sim 0.7$~keV and a Comptonizing medium of thermal electrons at $\sim 2$~keV, which is broadly consistent with our findings. \cite{yan2025} studied the 2023 outburst of \aql using Insight-HXMT observations and trace the transition from the hard to soft state. In the hard state, they find that a Comptonized disk plus blackbody model, as well as a Comptonized blackbody plus disk model can describe the spectra. However, in the soft state, the Comptonized component disappears and they find that the best model prescription is a combination of a disk ($\sim 1$~keV) and two blackbody components ($\sim 2.5$~keV and $\sim 1.3$~keV). This is similar to our model~1 fits, however we find evidence for a single $\sim 1.6$~keV blackbody with a temperature somewhere between the two blackbody fits of \cite{yan2025}. As our observations span a small part of the outburst, we do not find significant spectral variations. 

\subsubsection*{\fouru}
As in the case of \aql, we find that the two model prescriptions adequately describe the persistent emission spectra of \fouru. In model~1, the spectra is described with a $\sim 1$~keV disk, along with a $\sim 1.7$~keV blackbody component, whereas using model~3, $\sim 0.6$~keV disk seed photons, are Comptonized by the Corona. Similar to \aql, the corona is optically thick. The similarity of the inferred spectral parameters for these two sources imply that they are in similar phases of their respective outbursts.

\cite{armaspadilla2017} fitted the broadband data of \fouru from Suzaku during its 2010 outburst decay using a hybrid three-component model \citep{lin2007}, comprising of the emission components from the NS surface, accretion disk, and the Comptonization. Our results using model~1 (disk and blackbody temperatures) are consistent with those reported by them for the soft state spectra, with the important distinction that we do not find any systematic residuals to account for a third component. \cite{bhattacherjee2024} fitted SXT, LAXPC, and NICER data of 2016 and 2020 outbursts of \fouru, when the source is in the banana state, using \texttt{thcomp*diskbb} (same as our model~3) and found good fits to the spectra. They also found that a broad Gaussian feature is required to improve the residuals with energy fixed at 6.9~keV. We also found evidence of disk reflection, with the presence of an iron line with best fit energy $\sim 6.7$~keV, and $\sigma\sim 0.1-0.3$~keV.

\subsubsection*{Trends of spectral parameters with flux}
\begin{figure*}
\centering
\begin{subfigure}{0.49\textwidth}
\includegraphics[scale=0.82, trim=0cm 1cm 0cm 0cm, clip=true]{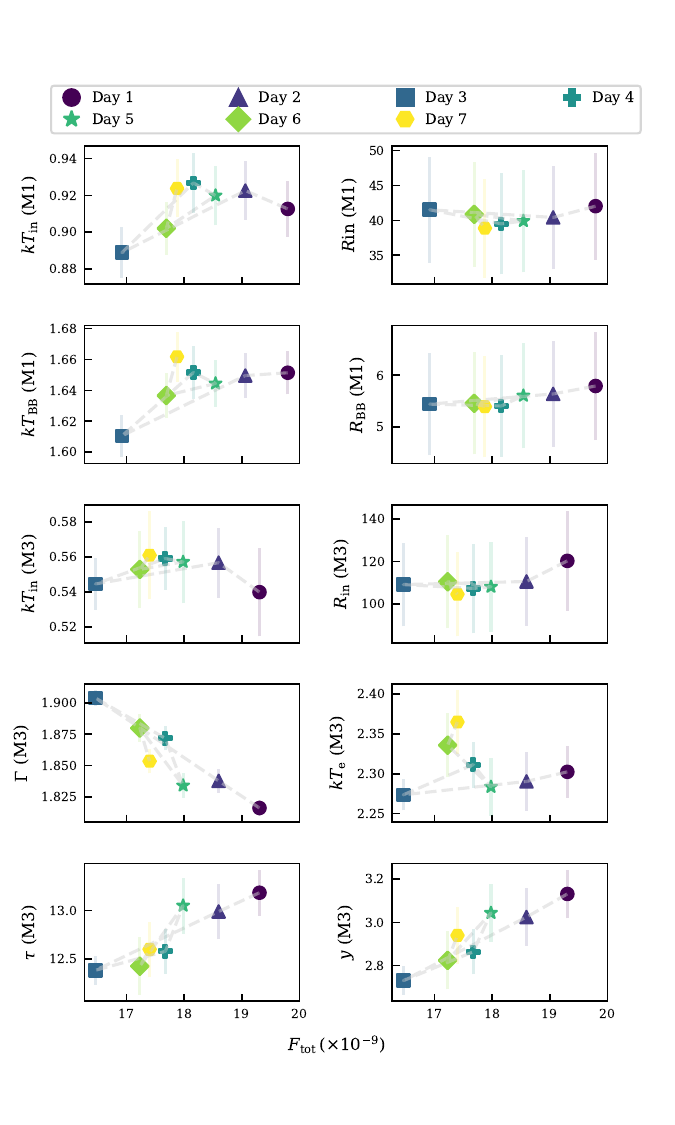}
\caption{}
\end{subfigure}\hfill
\begin{subfigure}{0.49\textwidth}
\includegraphics[scale=0.82, trim=0cm 1cm 0cm 0cm, clip=true]{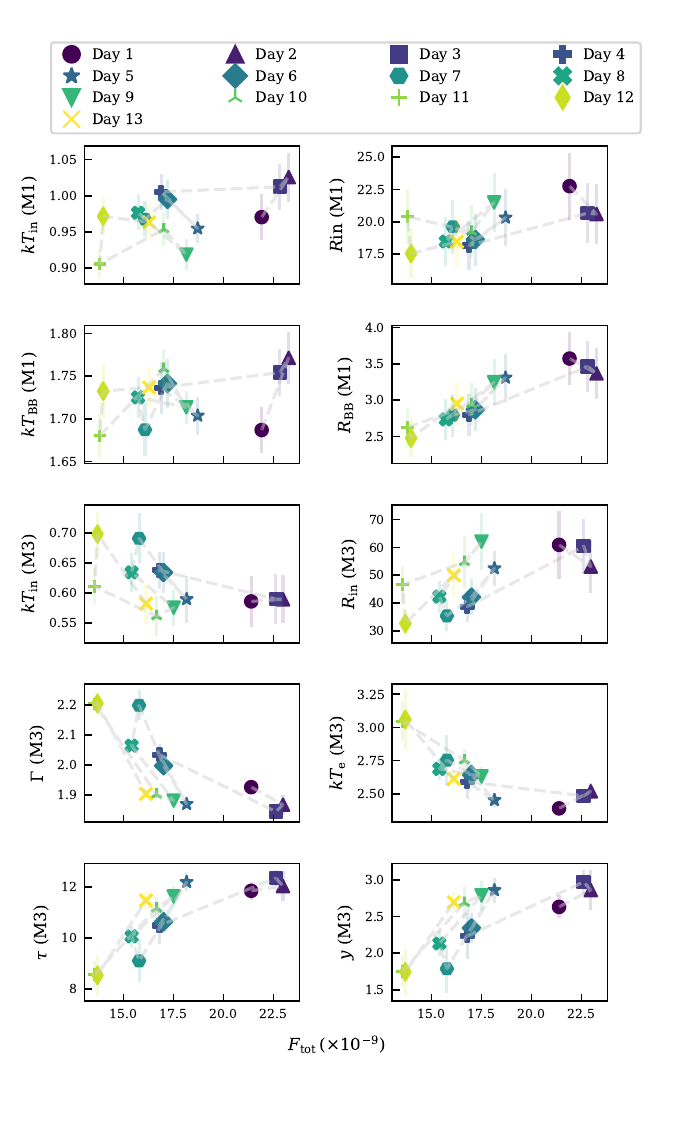}
\caption{}
\end{subfigure}
\caption{Variation of various spectral parameters of (a) \aql, and (b) \fouru, as a function of their total fluxes. Dashed lines are drawn to follow the evolution. `M1' and `M3' in the plot labels correspond to parameters derived from model~1 and model~3 respectively.}
\label{fig:paramswithflux}
\end{figure*} 

The observed persistent flux can be used as a proxy for the instantaneous accretion rate onto the neutron star, under the assumption that variations in luminosity primarily reflect changes in the mass inflow rate through the disk and boundary layer. The local accretion rate ($\dot{m}$) assuming uniform accretion over the surface of the neutron star is given by \citep{galloway2008}
$$\dot{m}=6.7\times 10^3\left(\frac{F_p}{10^{-9}\,\mathrm{erg}\,\mathrm{s}^{-1}\,\mathrm{cm}^{-2}}\right)\left(\frac{d}{\mathrm{10\,kpc}}\right)^2\left(\frac{M_{\mathrm{NS}}}{1.4 M_{\odot}}\right)^{-1}$$
\begin{equation}
\times \left(\frac{1+z}{1.31}\right)\left(\frac{R_{\mathrm{NS}}}{\mathrm{10\,km}}\right)^{-1}\,\mathrm{g}\,\mathrm{cm}^{-2}\,\mathrm{s}^{-1}
\end{equation}

During our observations, the persistent flux of \aql varied within the range $(1.6-2.0)\times 10^{-8}$~erg~s$^{-1}$~cm$^{-2}$, corresponding to $\dot{m} \simeq (2.7-3.4)\times 10^4$~g~cm$^{-2}$~s$^{-1}$ ($0.24-0.30$~$\dot{m}_{\mathrm{Edd}}$), while \fouru spanned $(1.3-2.3)\times 10^{-8}$~erg~s$^{-1}$~cm$^{-2}$, corresponding to $\dot{m} = (1.0-1.6)\times 10^4$~g~cm$^{-2}$~s$^{-1}$ ($0.08-0.14$~$\dot{m}_{\mathrm{Edd}}$). Both sources therefore remained within the soft (banana) branch of atoll sources throughout the observations, without undergoing large-scale spectral state transitions. 

Since the source flux does not show a monotonic trend with time but rather exhibits fluctuations on timescales of hours to days (Figures~\ref{fig:fullaql} and \ref{fig:full4u}), in this section we examine the variation of spectral parameters as a function of the total unabsorbed source flux. In Figure~\ref{fig:paramswithflux}, we plot the day-wise best fit spectral parameters of both the sources using models~1 and 3, as a function of the flux obtained with the respective models.

\begin{itemize}
\item Thermal components disk and blackbody (model~1): For both \aql and \fouru, model~1 reveals a weak increase in the disk temperature ($kT_{in}$) with increasing flux, while the inferred inner disk radius ($R_{in}$) remains approximately constant within uncertainties. In the limited accretion rate range explored here, an increase in accretion rate primarily leads to enhanced local dissipation and a higher effective disk temperature. The blackbody component, associated with emission from the neutron-star surface or boundary layer, broadly mirrors the behaviour of the disk temperature. For \fouru, unlike \aql, the blackbody radius ($R_{BB}$) exhibits a clear positive correlation with flux, which may indicate a increase in the effective emitting area as the accretion rate increases. 

\item Disk and Comptonized emission (model~3): When the spectrum is instead decomposed using a disk plus thermal Comptonization model (model~3), the inferred trends show notable differences, highlighting the model dependence of disk parameters in the presence of a Comptonizing medium. In \aql, $kT_{in}$ and $R_{BB}$ remains nearly constant with flux, 
while in \fouru a mild decreasing trend is observed is observed in $kT_{in}$, accompanied by an increase of $R_{in}$ with accretion rate. As the disk emission is modified by the Comptonizing region in this model, the difference in the inferred parameters could be due to a degeneracy between the disk and Comptonization parameters.

The Comptonization parameters show systematic variations with flux. For both sources, the optical depth of the Comptonizing region increases with increasing flux, accompanied by a corresponding increase in the Compton y-parameter. This indicates that, as the accretion rate rises, the Comptonizing medium becomes denser and Compton up-scattering of the soft photons by the coronal electrons becomes more efficient. Interestingly, for both sources we observe a hardening of the spectrum with increasing flux, reflected in a decrease of the photon index $\Gamma$. In addition, while the coronal temperature $kT_e$ does not vary much for \aql, likely due to the limited sampling of accretion rate, it shows a strong cooling trend with increasing fluxes \fouru. This harder-when-brighter behaviour can be attributed to enhanced Comptonization in the boundary layer or transition region. Our results suggest that, over the accretion-rate range probed here, the enhanced Comptonization efficiency outweighs the enhanced cooling by soft photons, leading to a net hardening of the spectrum.

\end{itemize}

\subsection{Superburst}
\label{subsec:disc_sb}
We have presented a detailed study of the superburst from \fouru observed by \xsp. We carried out time-resolved spectroscopy spanning the precursor (4UB1), superburst rise, and extended decay over the next several hours. We find that the spectra changes significantly with respect to the pre-burst (persistent) durations. Using model~3 (\texttt{thcomp*diskbb}) as our best-fit model of persistent emission along with an $f_a$ factor to allow for its scaling, we find that an additional blackbody component provides a satisfactory fit to the burst durations. The blackbody temperature rises to $\sim 2$~keV during the superburst peak, followed by a monotonic decline to $\sim 0.8$~keV over the next 5~orbits. This behaviour is consistent with thermal cooling of the neutron-star envelope following deep carbon ignition. Similar cooling behaviour has been observed in superbursts from 4U $1820-30$ and 4U~$1636-536$, where the burst spectrum is well described by a cooling blackbody over several hours (\citealt{strohmayer2002, kuulkers2004}).

During the entire superburst, we find that $f_a$ remains less than 1, contrary to what is observed during normal type-I bursts. This indicates a suppression of the persistent emission during the superburst, suggesting that intense radiation from the superburst may temporarily disrupt or reduce the inner accretion flow. This scenario is also supported by the fact that allowing the disk parameters to vary (from the pre-burst values) results in a higher inner disk radius with corresponding decrease in the inner disk temperature, as mentioned in Section~\ref{subsec:spec_sb}.  

A similar quenching of persistent emission was reported by \cite{peng2025} during the superburst of 4U~$1820-30$. They fitted a two component model (\texttt{compTT+bbodyrad}) to the superburst spectra and found that the persistent emission (Comptonization flux) is almost completely quenched during the superburst, which subsequently relaxes to pre-burst values over the next few hours. To see if \fouru shows a similar behaviour, we now let both components of persistent emission (model~3) freely vary, and fitted the six coarsely binned spectra covering the superburst. We find that the spectral shape changes significantly, and an additional blackbody component is statistically not required. Specifically, the spectra becomes considerably harder, the electron temperature is lower, accompanied by lower inner disk temperature, and higher disk radius, as compared to the persistent values (Table~\ref{tab:model3}). For example, in the first bin, the fit parameters are $\Gamma=1.26$, $kT_e=2.33$~keV, $kT_{in}=0.34$~keV, and $R_{in}\sim 259$~km, with $\chi^2_{\mathrm{red}}=1.09$. However, due to insufficient statistics, all parameters cannot be simultaneously constrained well, and moreover, this prescription is unphysical due to absence of blackbody component. 

We also note here that \cite{keek2014}, on the other hand, found that the persistent flux increases during the superburst and subsequently relaxes to the pre-superburst levels as the superburst decays. They use an absorbed cutoff powerlaw model to fit RXTE/PCA spectrum of 4U~$1636-536$ in the pre-superburst orbit, and add a blackbody component during the superburst. Keeping all other powerlaw parameters frozen, its normalization more than doubles during the superburst.

\cite{boztepe2025} study the effect of the 2020 superburst of \fouru on the accretion flow. They fit the NICER and HXMT-LE observations around the superburst using a combination of disk blackbody and blackbody emission, and find a systematic temporal evolution of the disk properties after the superburst, followed by a gradual recovery. We note here that \cite{boztepe2025} report an opposite trend in the disk behaviour, with the disk being hotter and closer to the NS than during persistent emission immediately after the 2020 superburst of \fouru, which subsequently shows cooling and recovery over the next few days. However, we do not see such trends over such long timescales. This apparent discrepancy could be due to the different energetics of the two outbursts -- \cite{boztepe2025} note that not only was the 2025 outburst `weaker' than the one in 2020 in terms of total fluence, peak outburst flux, and waiting time between the outburst beginning and superburst, the 2020 superburst also had a much higher ignition column depth than the 2025 one. Hence, these two superbursts may have different characteristics which cannot be directly compared. Other possible cause of the discrepancy could be in the different temporal sampling of the spectra in the two outbursts.

Using the full band light curve, we find that the e-folding time scale of the superburst is 3.05~h. However, the hard and soft components decay at different rates, with the hard band decaying faster as compared to the soft. The faster decay of the hard X-ray emission reflects the overall cooling of the neutron-star envelope following deep carbon ignition. As the effective temperature decreases with time, the blackbody peak shifts to lower energies, leading to energy-dependent decay timescales. This is also reflected in the initial increase in hard color near the superburst peak (Figure~\ref{fig:superburst}), and the subsequent dip below pre-burst levels suggests rapid cooling of these layers, consistent with the temperature evolution inferred from spectral fitting. A similar energy-dependent decay behaviour was observed by \cite{kuulkers2002} during the superburst of KS~$1731-260$, when the hard emission ($5-28$~keV) was found to decay faster than the soft ($2-5$ keV). 

As seen in many previous superbursts (e.g. \citealt{strohmayer2002, kuulkers2002, intzand2003}, see also \citealt{intzand2017}), we also observed a precursor $\sim 15$~minutes before the superburst onset. The rise and decay time scales of the precursor (4UB1) is comparable to those of the other type-I burst (4UB2) observed from this source (see Table~\ref{tab:burst_params}). However, the precursor has a significantly higher peak flux, temperature, and fluence. The $f_a$ value is also higher than the other burst. All these point to the fact that the precursor is more energetic than the normal type-I burst. For superbursts of some sources such as GX~$17+2$ \citep{intzand2004}, KS~$1731-260$ \citep{kuulkers2002}, and 4U~$1636-536$ \citep{kuulkers2004}, the precursor burst is either similar to or weaker than normal Type-I bursts from the same source. A notable exception is the case of the precursor of 4U~$1820-30$ \citep{strohmayer2002}, which shows hints of photospheric radius expansion (PRE). For the case of 4U~$1254-69$, \cite{intzand2003} find evidence of a double-peaked precursor burst, which is brighter than the typical type-I bursts observed from this source. Similar findings are also reported by \cite{strohmayer_markwardt2002} for 4U~$1636-536$.

The precursor burst in 4U~$1820-30$ occurs $\sim 13$~s before the superburst onset, whereas in the case of 4U~$1636-536$, the 2001 superburst actually started between $\sim 145$~s and $\sim 46$~min \textit{before} the precursor \citep{kuulkers2004}. In other cases such as KS~$1731-260$, the exact duration between the precursor and onset cannot be determined (estimated $\lesssim 13$~mins) owing to data gaps. Thus, properties of precursor bursts have been found to vary across different cases. However, it is to be noted that for many of the superbursts, the onset is not fully captured or the statistics are not sufficient for a confident detection and characterization of the precursor burst. As a result, our present knowledge of precursors and their relation to the observed superburst is incomplete.
 
Considering the total unabsorbed $0.8-15$~keV flux, the peak flux of the precursor burst reaches $\sim 1.8\times 10^{-7}$~erg~cm$^{-2}$~s$^{-1}$, which reaches the Eddington limit. As described in Section~\ref{subsec:tim_sb}, the rising phase of the superburst cannot be fitted with a single `fast rise', and has been modeled using three piecewise linear functions to follow the rise characteristics. The time duration between the precursor peak and superburst peak is $\sim 23$~minutes, with a `rising' time of $\sim 13.6$~mins following the slower-than-usual decay of the precursor. We estimated the timescale of thermal diffusion (\citealt{cumming2004, meisel2018}) following the Carbon ignition to explain the superburst rise as $t_{th} = H^2/D$, where $H$ is a characteristic length scale and $D$ is the thermal diffusivity. We have $H=y/\rho$ and $D=K/\rho C_P$, where $\rho$ is the density, $K$ is the thermal conductivity (dominated by electron-ion collisions), and $C_P$ is the heat capacity. Plugging in typical values (\citealt{potekhin1999, cumming2001, cumming2004}), we obtain diffusion timescales of several hours. However, this picture of the superburst rise may be too simplistic. As modeled by \citealt{weinberg2006, weinberg2007}, the superburst rise can be divided into three phases -- a convective stage, a local thermonuclear runaway, and a hydrodynamic stage. In the last phase, a combustion wave forms which can propagate either as a deflagration or a detonation (in this case, driving a shock wave into surrounding layers). Their model predicts the beginning of the superburst to be $\sim 100$~s after the ignition. Lightcurves produced by 1D multizone models of superbursts (\citealt{keek2011, keek2012}) also show the rising phase of the superburst to last from a few tens to hundreds of seconds, consistent with our observations.

\begin{figure*}
\centering
\begin{subfigure}{0.33\textwidth}
\centering
\includegraphics[scale=0.27]{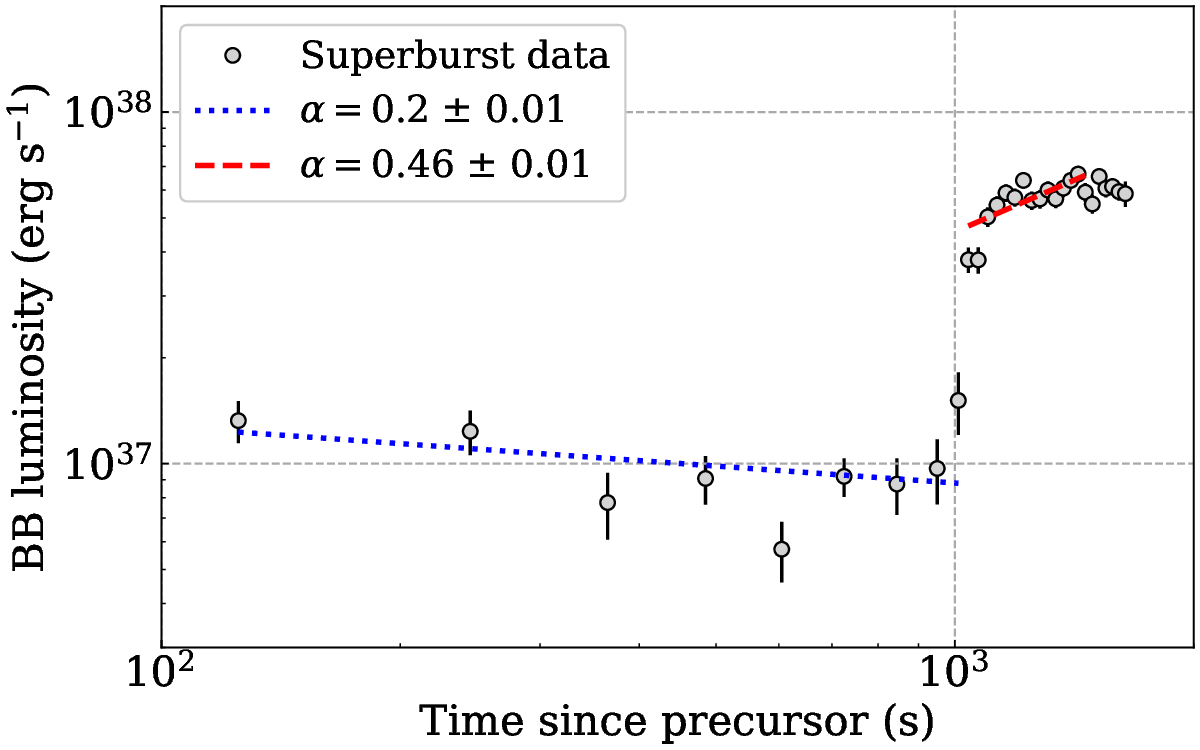}
\caption{}
\end{subfigure}
\begin{subfigure}{0.33\textwidth}
\centering
\includegraphics[scale=0.27]{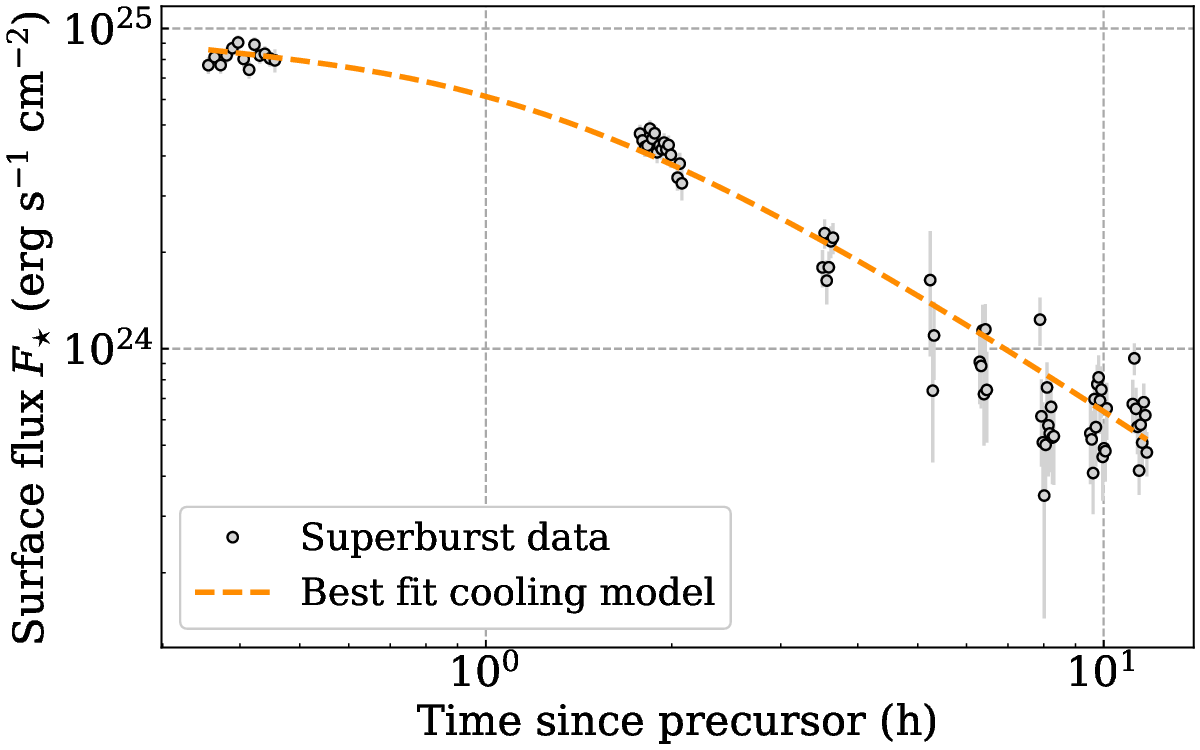}
\caption{}
\end{subfigure}
\begin{subfigure}{0.33\textwidth}
\centering
\includegraphics[scale=0.27]{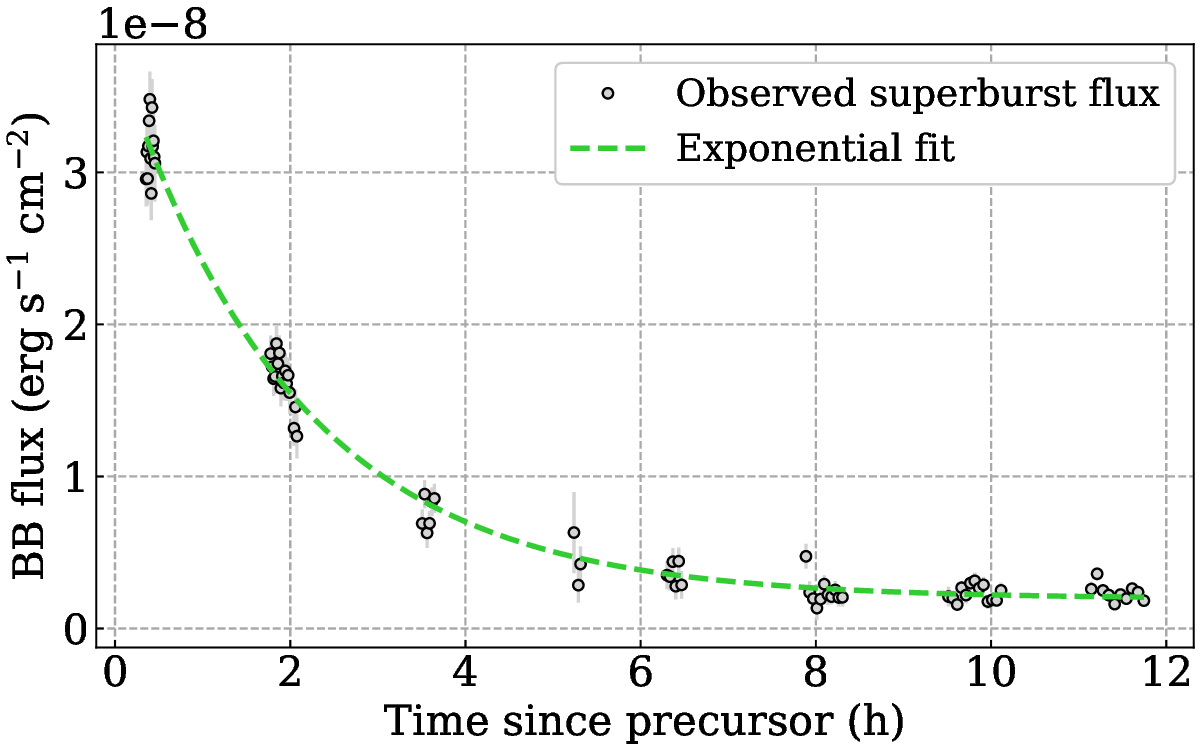}
\caption{}
\end{subfigure}  
\caption{(a) Fit to the blackbody luminosity evolution with time during the initial phase of the superburst using the prescription of \citet{keek2015}, (b) Evolution of blackbody surface flux decay after the superburst using the cooling model of \citet{cumming2004}, (c) Exponential fit to superburst flux decay for computing the net X-ray fluence of the event.}
\label{fig:lc_modeling}
\end{figure*}

\cite{keek2015} show that the shape of the rising part of the light curve encodes valuable information about the initial temperature profile of the NS following Carbon ignition. To test the same for our superburst observation, we compute the blackbody luminosity, corresponding to the superburst emission (over and above the persistent emission) as a function of time from the spectral fits described in Section~\ref{subsec:spec_sb}. Considering the precursor burst peak time as $t=0$, we fit the luminosity evolution between the precursor decay ($t\sim 60$~s) and superburst peak ($t\sim 1400$~s), as shown in Figure~\ref{fig:lc_modeling}a. From the shape of the light curve, two distinct slopes are apparent, corresponding to $\alpha=0.2$ and 0.5 in the initial ($t\sim  100-1000$~s) and later ($t\sim 1000-1500$~s) phases respectively. The $\alpha$ value indicates the shape of initial temperature profile with depth ($T\propto y^{\alpha}$), with $\alpha\sim 0.125$ indicating a local burning of fuel with no heat transport and $\alpha\lesssim 0.3$ for an adiabatic profile.

\cite{cumming2004} and \cite{cumming2006} provide a model of the thermal evolution of the NS surface layers after the burst, by fitting the evolution of the surface flux $F_{\star}$, where $F_{\star}=(1+z)^2\left(\frac{d}{R_{NS}}\right)^2F_{obs}$. Their model is parametrized by the energy released per gram $E_{17}$ ($\times 10^{17}$~erg~g$^{-1}$) and ignition column depth $y_{12}$ ($\times 10^{12}$~g~cm$^{-2}$). We fitted the blackbody flux decay during the superburst using this model (Figure~\ref{fig:lc_modeling}b) and found $E_{17}=2.04$ and $y_{12}=1.48$, consistent with values reported for previous superbursts from this source (\citealt{keek2008, boztepe2025}), as well as from other sources (e.g. \citealt{serino2016}). This value of $E_{17}$ corresponds to a mass fraction of $\sim 20\%$ of burnt carbon, consistent with values obtained for other sources. The fitted $E_{17}$ and $y_{12}$ values imply a total energy release of $3.8\times 10^{42}$~erg.

To estimate the energy released in the form of X-ray radiation, we fitted the exponentially decaying blackbody flux during the superburst, and integrated it over $\sim 11.4$~h as shown in Figure~\ref{fig:lc_modeling}c. We find the net X-ray fluence to be $\sim 3\times 10^{-4}$~erg~cm$^{-2}$, which corresponds to a total energy of $6.3\times 10^{41}$~erg at the surface of NS, after correcting for the gravitational redshift. This is only $\sim 16\%$ of the inferred energy release from the cooling models, which suggests that the rest of the energy may be released via other avenues such as neutrino losses, conduction to deeper layers (which are radiated on much longer timescales), etc. Moreover, the X-ray energy released during the superburst is $\sim 350$ times higher than that released during normal type-I bursts from the source ($E_{rad}$ in Table~\ref{tab:burst_params}). The $y_{ign}$ inferred from the X-ray fluence (equation~5) using a $Q_{\mathrm{nuc}}$ value of 0.1~MeV~nucleon$^{-1}$ is $\sim 4\times 10^{11}$~g~cm$^{-2}$, a factor of $3-4$ smaller than that found from cooling models.

We found an e-folding time scale of $\sim 3.1$~h by fitting the superburst decay light curve. Another measure of the timescale can be estimated from $\tau=E_b/F_{\mathrm{peak}}$ \citep{peng2025}, where $E_b$ is the burst fluence and $F_{\mathrm{peak}}$ is the peak superburst flux. From spectral fitting (Section~\ref{subsec:spec_sb}), we found the peak flux to be $3.4\times 10^{-8}$~erg~s$^{-1}$~cm$^{-2}$ considering the flux of only the blackbody component. This gives $\tau$ as $\sim 2.4$~h, slightly lower than the estimated value from light curve.

A well-observed phenomenon is the cessation of normal type-I X-ray bursts for a few days to weeks following a superburst (see e.g. \citealt{cornelisse2000, kuulkers2002, intzand2004, kuulkers2004}). We observe a type-I X-ray burst from the source just $\sim 5.4$~days after the superburst (calculated from the superburst peak). This is an order of magnitude lower than the previously reported quenching times from this source: 99.8~days in 2005 \citep{keek2008} and 58.98~days in 2020 \citep{boztepe2025}. Note that due to gaps in the observation, all these numbers correspond to upper limits. \cite{cumming2001} propose that the burst quenching can be explained by the stabilization of H/He burning owing to the influx of heat from the deeper layers (where the superburst ignites) to the H/He burning layers. The predictions from the cooling model by \cite{cumming2004} match well with the observed quenching time scales, where the resumption of type-I bursts marks the flux at these layers falling below a critical `stabilizing flux' as the superburst cools. 

\section{Summary and Conclusions}
We have presented a detailed spectro-temporal analysis of the \xsp observations of two sources, \aql and \fouru. The observations are during the initial phases of the respective outburst decays, when the source is in the soft state. We find the best fits to the persistent emission spectra from the sources using two alternate model combinations: a single temperature blackbody along with disk blackbody (model~1), and a thermally Comptonized disk blackbody model (model~3). We study the variation of spectral parameters during the observations and find systematic variations with flux. In particular, we find that the optical depth of the Comptonization component increases at higher accretion rates, accompanied by a hardening of the spectra. 

We also study the superburst from \fouru, and the type-I bursts observed during the observations using time-resolved spectroscopy. The persistent emission model scaled by a constant factor, along with a blackbody component provide good fits to the burst and superburst spectra, with a cooling blackbody component as they decay. Interestingly, contrary to the normal type-I bursts, the persistent emission is found to be suppressed during the superburst. The superburst is ignited at much greater depths, and release hundreds of times more energy as compared to the type-I bursts. 

\section*{Acknowledgements}
The XPoSat project is managed and facilitated by Indian Space Research Organisation. The authors are grateful to the anonymous referee for their valuable comments and suggestions, which helped improve the quality of this manuscript. The authors thank XPoSat project team, facilities team, assembly, integration, and checkout teams, and mission team for their involvement and support in enabling XSPECT payload on XPoSat mission. The authors are grateful to GD, SAG; DD, PDMSA, and Director, URSC for continuous encouragement and support to carry out this research. 

\section*{Data Availability}
The data utilized in this study are hosted at the Indian Space Science Data Centre (ISSDC) and available for download via PRADAN (https://pradan1.issdc.gov.in/x01).



\bibliographystyle{mnras}
\bibliography{references}

\bsp	
\label{lastpage}
\end{document}